\documentclass[runningheads]{llncs}
\usepackage[T1]{fontenc}
\usepackage{graphicx}
\usepackage{amsfonts,amssymb,bm}
\usepackage{diagbox}
\usepackage{marvosym}
\usepackage{xspace}
\usepackage{pifont}
\usepackage{multirow}
\usepackage{comment}
\usepackage{amsmath}
\usepackage{booktabs}

\usepackage{enumitem}

\usepackage{tcolorbox}

\newcommand{\attackname}{\textsc{LaserGuider}\xspace}

\usepackage{verbatim}
\usepackage{hyperref}
\usepackage{balance}

\usepackage{colortbl}

\usepackage{flexisym}

\newcommand{\red}[1]{\textcolor[rgb]{0.85, 0, 0}{#1}}

\usepackage{cleveref}

\usepackage{makecell}

\usepackage[flushleft]{threeparttable}

\usepackage{algorithm}
\usepackage[noend]{algorithmic}

\algsetup{indent=18pt}

\usepackage[switch]{lineno}

\usepackage{cite}

\usepackage{subfig}

\usepackage{fancyhdr}

\usepackage[flushleft]{threeparttable}

\usepackage{etoolbox}
\newcommand{\samethanks}[1][\value{footnote}]{\footnotemark[#1]}

\begin{document}
\title{\attackname: A Laser Based Physical Backdoor Attack against Deep Neural Networks\thanks{In Proceedings of the 23rd International Conference on Applied Cryptography and Network Security (ACNS), Munich, Germany, 23-26 June, 2025}}


\author{Yongjie Xu\thanks{Co-first authors \& equal contribution,
where the order is determined by a random coin flip.}\inst{1}\orcidID{0000-0001-5816-8031} \and
Guangke Chen\samethanks\inst{2}\orcidID{0000-0001-8277-3119} \and \\ 
Fu Song\textsuperscript{\Letter}\inst{3,4}\orcidID{0000-0002-0581-2679} \and 
Yuqi Chen\inst{1}\orcidID{0000-0003-2988-6012} 
}
\authorrunning{Y. Xu, G. Chen, et al.}
%
\institute{ShanghaiTech University \email{\{xuyj,chenyq\}@shanghaitech.edu.cn} \and 
Pengcheng Laboratory \email{chengk@pcl.ac.cn} \and
Key Laboratory of System Software (Chinese Academy of Sciences) and State Key Laboratory of Computer Science, Institute of Software, Chinese Academy of Science \and 
Nanjing Institute of Software Technology \email{songfu@ios.ac.cn}}

\maketitle              

\begin{abstract}
Backdoor attacks embed hidden associations between triggers and targets in deep neural networks (DNNs), causing them to predict the target when a trigger is present while maintaining normal behavior otherwise. Physical backdoor attacks, which use physical objects as triggers, are feasible but lack remote control, temporal stealthiness, flexibility, and mobility. 
 To overcome these limitations, 
      in this work, we propose a new type of backdoor triggers utilizing lasers that feature long-distance transmission and instant-imaging properties. Based on the laser-based backdoor triggers, we present a physical backdoor attack, called \attackname,
      which possesses remote control ability and achieves high temporal stealthiness, flexibility, and mobility. 
      We also introduce a systematic approach to optimize laser parameters for improving attack effectiveness. 
      Our evaluation on traffic sign recognition DNNs, critical in autonomous vehicles, demonstrates that \attackname with three different laser-based triggers achieves over 90\% attack success rate with negligible impact on normal inputs. Additionally, we release LaserMark, the first dataset of real-world traffic signs stamped with physical laser spots, to support further research in backdoor attacks and defenses.
\end{abstract}

\section{Introduction}
\label{sec:intro}
Recently, deep neural networks (DNNs) have been widely deployed in various applications such as autonomous driving~\cite{autonomous-driving-survey}, medical diagnosis~\cite{medical-diagnosis-survey}, media understanding~\cite{speech-survey, IJCS_video}, and attack detection~\cite{IJCS_dos}. While their success is attributed to increasingly complex structures and large capacities, DNNs require substantial data and computation for training. This has led to a growing reliance on third-party datasets, platforms, or pre-trained DNNs~\cite{li2022backdoor}, introducing vulnerabilities to backdoor attacks~\cite{gu2017badnets, backdoor_detection, turner2019label, chen2017targeted, liu2017neural, chen2023feature, lian2023ipcadp, IJCS_multi_backdoor, backdoor_free_defense,ZhaoCWYS021}. 
Backdoor attacks aim to embed hidden associations between triggers and targets into DNNs during training. During inference, DNNs misclassify inputs with triggers to attacker-specified targets while performing normally on inputs without triggers.

In reality, DNNs are invocated as follows: physical objects are captured as analog signals by sensors, converted to digital inputs, and then fed into DNNs for prediction. Early digital backdoor attacks assume direct manipulation of digital inputs~\cite{gu2017badnets, turner2019label, chen2017targeted, liu2020reflection, zhong2020backdoor, liu2017trojaning, composite-attack, input-aware-attack, LIRA, backdoor-NLP, speaker-backdoor}, making them less effective or infeasible in real-world scenarios. Later attacks adopt physical triggers to manipulate physical objects to activate backdoors~\cite{gu2017badnets, chen2017targeted, wenger2021backdoor, li-backdoor-physical-world}, improving practicality. However, current physical backdoor attacks face limitations: lack of remote control ability, low temporal stealthiness, and low flexibility and mobility (see \Cref{sec:trigger-design}).

We observe that these limitations are attributed to that their adopted physical triggers 
(e.g., square stickers~\cite{gu2017badnets}, sunglasses~\cite{chen2017targeted}, and earrings~\cite{wenger2021backdoor}) 
have to be physically attached onto the physical objects. 
This motivates us to design a new type of physical triggers. 
After investigation, we choose laser spots as physical triggers, 
due to its notable advantages as follows: 
(1) Laser possesses the long-distance transmission ability, thus can be projected to the physical objects remotely. 
(2) Laser possesses the instant-imaging capability, 
thus the adversary can turn on a laser pointer to project the laser  onto a physical object \emph{right before} the physical object got captured by sensors, achieving high temporal stealthiness. 
(3) The projection position can be instantaneously switched from one physical object 
to another object, achieving high flexibility and mobility 
during a consecutive attack. 

Based on our novel laser-based triggers, we propose a physical backdoor attack, named \attackname. 
\attackname embeds backdoors by poisoning training datasets. To reduce the cost of collecting training samples with physical laser spots, we simulate physical triggers using digital laser-based triggers, which are used to poison clean digital samples. 
We also propose an effective approach to optimize these digital triggers with adjustable parameters to achieve more powerful physical backdoor attacks. 

We evaluate \attackname on state-of-the-art traffic sign recognition DNNs commonly used in autonomous driving~\cite{gu2017badnets, duan2021adversarial}. Experiments on a self-collected real-world dataset with physical laser spots confirm that \attackname achieves 90.5\%, 93.2\%, and 95.3\% attack success rates with three types of laser-based triggers (differing in color and shape) while negligibly impacting normal inputs. Additionally, \attackname supports many-to-one and many-to-many backdoor attacks, allowing multiple triggers to activate the same or distinct backdoors. An ablation study further validates the effectiveness and generalizability of our laser parameter optimization approach.

In summary, the main contributions of this work are: 
\begin{itemize}
  \item We design a new type of laser-based backdoor triggers based on which 
 we propose a physical backdoor attack \attackname,  featuring remote control ability, 
  high temporal stealthiness, and high flexibility and mobility, thanks to the unique properties of lasers.  
  \item We propose a systematic approach to  optimize the parameters of laser-based triggers,
   enhancing the attack success rate of our physical backdoor attack.
  \item We capture, collect, and clean traffic sign images with physical laser-based triggers in the real-world, 
  and release them as the first public dataset LaserMark to facilitate further research in this field.
\end{itemize}

\smallskip
\noindent
{\bf Organization.} \Cref{sec:background-related} covers background and related work on backdoor attacks. \Cref{sec:methodology} introduces our backdoor attack \attackname, including the threat model (\Cref{sec:threat-model}), overview (\Cref{sec:overview}), laser-based trigger design (\Cref{sec:trigger-design}), and laser parameters optimization approach (\Cref{sec:opt}). \Cref{sec:setting} details the experimental setup, especially our publicly available dataset, LaserMark. Evaluation results are presented in \Cref{sec:exper-result}. \Cref{sec:discussion} discusses failure cases and potential countermeasures. \Cref{sec:concl} finally concludes the work. 

\smallskip \noindent {\bf Acronyms.} 
We summarize in \tablename~\ref{tab:notation} the acronyms used in this work. 
Our code and dataset are available at \url{https://github.com/S3L-official/LaserGuider}.

\begin{table}[h]
  \caption{Main acronyms.}
  \centering
 {
 \setlength{\tabcolsep}{2pt}
  \begin{tabular}{c|c}
    \hline
    Acronym & Description   \\ \hline
    \textbf{$T_{train}$} & clean training dataset   \\ \hline
     \textbf{$T_{test}$} & clean test dataset \\ \hline
  \textbf{$P_{train}$} & {p}oisoned training dataset with digital triggers  \\ \hline
   \textbf{$P_{test}$} & {p}oisoned test dataset with physical triggers \\ \hline
   \textbf{$T_{select}; {\alpha}$} & selected subset of $T_{train}$ for adding digital triggers; ${\alpha=\frac{|T_{select}|}{|T_{train}|}}$ \\ \hline
   {\textbf{$x; \delta; x^\delta; y; y^t$}} &  {image; trigger; image with trigger; label; target label} \\ \hline
  \textbf{$A_{p}/{A_{pn}}$} & 
  \makecell[c]{portion of images in $P_{test}$ which are classified as\\the target label {on poisoned/normal models}} \\ \hline
   \textbf{$A_{c}/{A_{cn}}$} &  \makecell[c]{portion of correctly classified images in $T_{test}$\\{on poisoned/normal models}} \\ \hline
  \end{tabular}
  }
  \label{tab:notation}
\end{table}

\smallskip \noindent {\bf Ethical Considerations.} 
(1) {\it Strictly Controlled Experiments:} We targeted open-source models and ensured our attack evaluation caused no real-world harm. Poisoned datasets and models were not released online. 
(2) {\it Discussion of Countermeasures:} We discuss countermeasures in \Cref{sec:defense} to mitigate potential threats, offering insights for stronger defenses.

\section{Background \& Related Works}\label{sec:background-related}
\subsection{Backdoor Attacks}
Backdoor attacks on DNNs insert secret associations between
triggers and targets into DNNs  
so that they perform normally on inputs without triggers, 
but behave maliciously as the adversary intended 
in the presence of a designated trigger. 
The feasibility of such attacks lies in the loss of control over training and/or training datasets, 
e.g., adopting third-party datasets instead of collecting and annotating own training dataset, 
using third-party training platforms instead of training locally, 
and even directly deploying third-party pre-trained DNNs rather than training from scratch~\cite{li2022backdoor}.

Backdoor can be embedded in various different ways, such as poisoning training datasets~\cite{gu2017badnets, wenger2021backdoor}, 
modifying the model parameters~\cite{rakin2020tbt}, or adding extra sub-modules~\cite{tang2020embarrassingly}, 
among which training dataset poisoning is the most typical and widely adopted strategy~\cite{li2022backdoor}.
Specifically, the adversary creates a poisoned training dataset by 
mixing normal samples with poisoned samples, 
where each poisoned sample consists of 
a data point with a designated trigger and a target label chosen by the adversary. Training on such poisoned dataset results in a backdoored DNN with the adversary-chosen
behavior in the presence of triggers.
During inference, the adversary attaches a trigger to the input sample
to activate the backdoor, forcing the DNN to behave maliciously as intended. 

\subsection{Related Works}\label{sec:related_work}
Current backdoor attacks can be broadly categorized into digital backdoor attacks and physical backdoor attacks. 

\smallskip \noindent {\bf Digital backdoor attacks.} 
Such attacks~\cite{gu2017badnets, turner2019label, chen2017targeted, liu2020reflection, zhong2020backdoor, liu2017trojaning, composite-attack, input-aware-attack, LIRA, backdoor-NLP, speaker-backdoor,Sun0SN022,Sun0S023} assume that the adversary can manipulate DNNs' digital inputs to insert triggers. While well-explored, this assumption limits their real-world applicability, where physical inputs are captured by sensors and converted to digital form, preventing direct manipulation. {This motivates our focus on physical backdoor attacks.}

\smallskip
\noindent {\bf Physical backdoor attacks.} 
In physical attacks, the adversary manipulates physical inputs to influence digital ones, using objects as backdoor triggers, such as square stickers~\cite{gu2017badnets}, sunglasses~\cite{chen2017targeted}, and earrings~\cite{wenger2021backdoor}. These triggers activate the backdoor during inference by attaching them to physical objects, like placing a sticker on a traffic sign for traffic sign recognition DNNs~\cite{gu2017badnets} or wearing sunglasses for face recognition DNNs~\cite{wenger2021backdoor}. While more practical than digital attacks, current physical backdoor methods suffer from limitations in remote control, temporal stealthiness, flexibility, and mobility. 
We propose a laser-based physical backdoor attack to address these limitations.

\smallskip \noindent {\bf Attacks based on lighting \& laser.}
In \cite{li2020light}, 
an LED-based color stripe pattern is utilized as a backdoor trigger. The LED parameters (e.g., intensity) are optimized using evolutionary computing based on the target DNN's predictions. Their attack targets the unique registration phase of face recognition by embedding patterns into the victim's registered face and the adversary's face captured by cameras, causing the DNN to misidentify the adversary as the victim. In contrast, our attack impacts the training phase by poisoning training datasets, offering a broader model and application scenario generalizability.
In addition, their attack still lacks remote control ability and suffers from low feasibility 
since the LED needs to be placed very close to the physical objects and the cameras,
and has to be placed ahead over each physical object to be attacked. 

Another related work, AdvLB~\cite{duan2021adversarial}, projects laser beams onto physical objects to create adversarial examples that fool DNNs including autonomous driving. AdvLB optimizes laser parameters using greedy methods. The key differences are: (1) AdvLB focuses on inference-phase adversarial attacks, while \attackname targets training-phase backdoor attacks; and (2) AdvLB uses laser beams, whereas \attackname employs laser spots, involving distinct laser parameters.
\section{Methodology of \attackname}\label{sec:methodology}

\subsection{Threat Model}\label{sec:threat-model}\label{sec:model}
\noindent {\bf Adversary's goal.} 
{Backdoor attacks aim to make victim DNNs behave as intended by adversaries when triggers are present, with varying goals depending on DNNs' application scenarios, such as traffic sign recognition~\cite{gu2017badnets, duan2021adversarial}, face recognition~\cite{wenger2021backdoor, li2020light}, person re-identification~\cite{person_re_identification_backdoor_attack}, lane detection~\cite{Lane_Detection_backdoor, Lane_Detection_backdoor_2}, OCR~\cite{OCR_backdoor}, and pedestrian detection~\cite{person_re_identification_backdoor_attack}. For example, in traffic sign recognition, a stop sign with a trigger may be misclassified as a speed limit sign, causing potential car collisions. 
In face recognition, the face of imposters (resp. victims) with triggers may be recognized as from victims (resp. imposters), causing authentication bypass (resp. DoS).
\attackname, is general and applicable across scenarios. Following prior works~\cite{gu2017badnets, duan2021adversarial, tf_1, tf_2, tf_3, tf_4, tf_5, tf_6, tf_7, tf_8}, we evaluate it using traffic sign recognition as a case study, a critical and widely used component in autonomous driving. While adversaries may lack direct financial motivation to target self-driving cars~\cite{Robustness}, traffic sign recognition adequately evaluates the attack.}

Meanwhile, the adversary intends to achieve stealthiness and mobility. 
The stealthiness is two-fold: (1) the victim DNNs should maintain sufficient performance 
on normal inputs without triggers, otherwise they will be discarded; 
(2) the trigger adding and removal should leave attack evidence 
as little as possible. 
Existing physical backdoor attacks typically necessitate close proximity to the targeted physical objects 
and sufficient time slots before and after the attack to add and remove the triggers, respectively,
significantly heightening the exposure risk. 
{Mobility means triggers can be added conveniently and efficiently to multiple physical objects for consecutive and real-time attacks, unlike existing methods requiring the pre-deployment of triggers to multiple traffic signs.}

\smallskip
\noindent {\bf Adversary's capability.}
Firstly, we assume that the adversary can manipulate a dataset and upload the resulting dataset online, 
which will be downloaded and used for training DNNs by developers.  
{However, the adversary cannot \emph{modify} any other aspects of training such as DNN architecture and training algorithm. }
Accordingly, the adversary chooses to insert the backdoor by poisoning training datasets 
instead of modifying DNN parameters or adding malicious sub-modules. 
{Despite this, the adversary may \emph{know} the details of target DNNs. 
This is reasonable considering the wide adoption of public well-known model architectures and typical training algorithms (e.g., Adam~\cite{ADAM}).}
Secondly, the adversary cannot directly modify the digital inputs when launching attacks at inference time, 
so needs to launch physical backdoor attacks by manipulating the physical objects. 
Regarding the physical attack, we assume that the adversary cannot touch the physical objects 
to add or remove any triggers, e.g., many traffic signs are erected in high places, 
{the adversary cannot touch the face of victims for causing DoS.} 
Despite this, the adversary can maintain a reasonable distance from the physical objects 
such that the physical objects can be recognized by the adversary
and projected by common lasers.

\subsection{Overview of \attackname}\label{sec:overview}
\label{sec:attack}

\begin{figure}[t]
    \centering
        \includegraphics[width=0.98\textwidth]{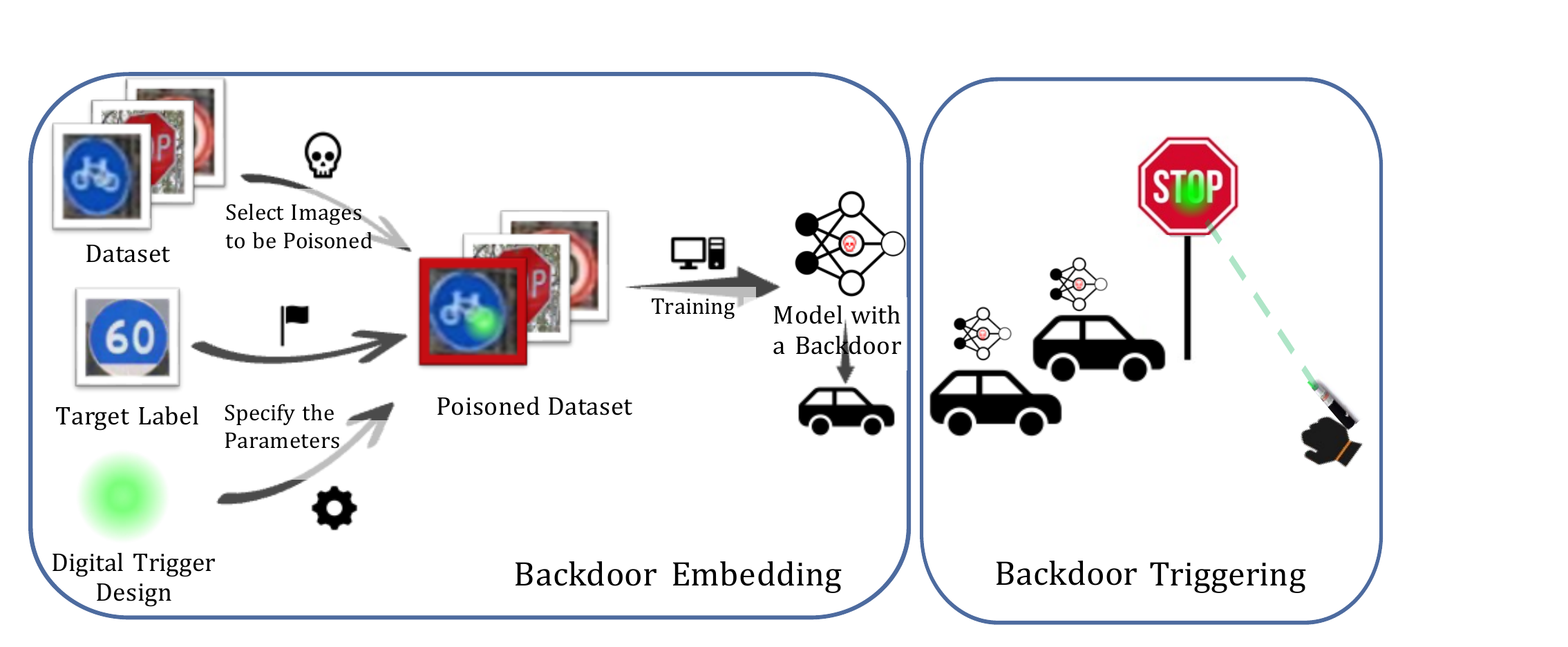}
        \caption{Overview of \attackname.}
        \label{fig:flow}
\end{figure}

The overview of our attack \attackname is depicted in \figurename~\ref{fig:flow}, 
consisting of two stages, namely, backdoor embedding and backdoor triggering. 

\smallskip
\noindent {\bf Backdoor embedding.} 
This stage injects a backdoor into DNNs by poisoning the training dataset,   
following the common practice of previous backdoor attacks~\cite{gu2017badnets, turner2019label, 
chen2017targeted, liu2020reflection, zhong2020backdoor, liu2017trojaning, composite-attack,input-aware-attack, LIRA}. 
First, the adversary specifies a target label $y^t$ 
and designs a \emph{digital} laser-based trigger $\delta$ (cf.~\Cref{sec:trigger-design}). 
Then, he selects a subset 
${T_{select}}$
of images from a legitimate training dataset 
${T_{train}}=\{(x_i,y_i)\}_{i=1}^{N}$ 
where 
$\alpha=\frac{|{T_{select}}|}{|{T_{train}}|}$
is a small ratio,  
and adds the trigger $\delta$ to each image $x$ in the chosen subset 
${T_{select}}$ 
and associates each modified image $x^\delta$ with the target label $y^t$, 
obtaining the poisoned training dataset 
${P_{train}=T_{train} \cup \{(x^{\delta}, y^t)|x\in T_{select}\}}$
Finally, the adversary publishes the poisoned dataset on the Internet, 
waiting for developers to download the dataset for training their DNNs, 
during which the backdoor will be naturally embedded into DNN models 
as a result of training with the poisoned dataset. 
{The dataset poisoning scheme enables our attack to process inherent model generalizability, 
i.e., the poisoned training dataset can be used by and affect any model. 
We will test our attack on four different models in \Cref{sec:exper-result}. 
}
Compared with previous backdoor attacks, \attackname features the laser-based trigger, 
the design of which will be elaborated in \Cref{sec:trigger-design}.

\smallskip
\noindent{\bf Backdoor triggering.} 
This stage activates the embedded backdoor in victim DNN models by adding \emph{physical} laser-based triggers to physical objects. 
To achieve this goal,  
the adversary simply turns on a laser pointer and projects the laser  onto the targeted physical object, e.g., traffic sign in this work, to which the autonomous vehicle equipped with the backdoored DNN model is going to photograph, 
forcing it to make a wrong prediction. 
To terminate the attack and destroy the attack evidences for being noticed, 
the adversary only needs to turn off the laser pointer. 
Moreover, during attack, the adversary is free to change the irradiation angle 
and orientation of the laser pointer, so that the laser spot can be quickly transferred 
and switched on the nearby traffic signs to achieve real-time and consecutive attacks, 
which significantly improve the attack efficiency considering the high mobility of autonomous vehicles. 
All these operations can be performed remotely with a reasonable distance from traffic signs and autonomous vehicles.

\subsection{Trigger Design}\label{sec:trigger-design} 
\noindent {\bf Design motivation.} 
Existing physical backdoor attacks utilize various physical objects as triggers, 
e.g., square~\cite{gu2017badnets}, earrings~\cite{wenger2021backdoor}, 
sunglasses~\cite{chen2017targeted}, bandana~\cite{wenger2021backdoor}, and tattoo~\cite{wenger2021backdoor}. 
These triggers are typically deployed in a ``sticker-pasting'' setting~\cite{duan2021adversarial}, 
where the adversary prints the triggers as stickers and affixes them onto physical objects.
Because of the ``sticker-pasting'' setting, these backdoor attacks suffer from the following three limitations:
\begin{itemize}
      \item {\bf Lack of remote control ability}.
      To control (e.g., adding or removing) triggers, 
      physical objects are required to be physically accessible to the adversary 
      so that the adversary can move close enough to touch and manipulate them.  
      This is inconvenient and is not always possible in practice, 
      e.g., many road traffic signs are erected on high places. 
      {e.g., adversaries cannot touch victims' face for causing DoS.}
      Also, the physical presence of the adversary increases the 
      risk of being noticed by victims and defenders.
      \item {\bf Low temporal stealthiness}. 
      {The trigger must be pasted to the targeted physical object in advance. 
      This typically leaves a sufficient time gap before the sensor captures the targeted physical objects~\cite{duan2021adversarial}, definitely increasing the risks of attacks being detected and subsequently removed.}
      \item {\bf Low flexibility and mobility}. 
      {The adversary attempts to achieve a consecutive attack where the victim DNN will be affected by multiple physical objects or the same physical object with different triggers. 
      Existing attacks require multiple triggers to be added \emph{in advance} into the physical objects that are likely to be captured, reducing flexibility, mobility, and efficiency.}
\end{itemize}

{We remark that these limitations are general across DNNs' application scenarios (at least for those in \Cref{sec:threat-model}). For face recognition, they pose no issue for causing authentication bypass but are crucial for causing DoS. Even LED-based triggers~\cite{li2020light}, which avoid ``sticker-pasting'', 
still suffers from these limitations (cf.~\Cref{sec:related_work}). 
}

This work is motivated by the following question: 
``{\it how to design a backdoor trigger to enable a backdoor attack with better stealthiness, convenience, efficiency, flexibility, and mobility in the physical real-world?}''

\smallskip
\noindent {\bf Design.}
After investigation, in this work, we propose to use laser spots as physical triggers due to their unique and desirable properties as follows.

Firstly, lasers have high brightness and exhibit highly collimated propagation without significant divergence, 
hence can maintain high power density even after a long-distance transmission. 
As a result, the adversary can manipulate the targeted physical objects by projecting lasers onto them from a long distance, 
\emph{achieving remote control ability.}

Secondly, 
as the swiftest entity in the world, 
lasers possess the instant-imaging property, i.e., 
there is a very short delay between the moment the laser is emitted 
from the laser pointer and when it is projected onto the intended targeted physical object. 
Hence, the adversary can launch the attack in a blink 
right before the targeted physical object is captured by the imaging sensor 
by simply turning on a laser pointer emitting the laser, 
rather than deploying the attack in advance, \emph{achieving high temporal stealthiness.} 
Also, 
to launch consecutive attacks, the adversary can simply move the projection of lasers 
{from one physical object to nearby ones}
by simply switching the laser pointer's projection position or using multiple laser pointers, 
\emph{achieving high flexibility and mobility.}

Lastly, lasers exhibit exceptional anti-interference capabilities and formidable penetrating power. 
As a result, 
the triggers and attacks possess a remarkable resilience 
against environmental disturbances, 
such as intense light, rainfall, and fog. This ensures their unwavering effectiveness even in challenging conditions.

\smallskip
\noindent {\bf Digital \& physical laser-based triggers.}
To obtain the poisoned training data, 
the adversary can photograph 
{physical objects}
projected with physical laser-based triggers in the real world. 
However, this involves substantial labor effort.
Therefore, when constructing a poisoned training dataset, 
we use digital laser-based triggers to simulate the patterns of physical triggers. Specifically, given the specified laser spot color and shape, 
it is easy to create a digital trigger that fulfills the requirements through geometric drawing.

\begin{figure}[h]
      \centering
      \subfloat[Triggers]{
      \begin{minipage}[t]{0.25\textwidth}
            \centering
            \includegraphics[width=0.4\textwidth]{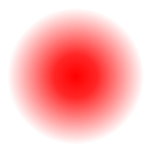}
      \end{minipage}\quad
      \begin{minipage}[t]{0.25\textwidth}
            \centering
            \includegraphics[width=0.4\textwidth]{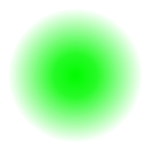}
      \end{minipage}\quad
      \begin{minipage}[t]{0.25\textwidth}
            \centering
            \includegraphics[width=1\textwidth, height=0.4\textwidth]{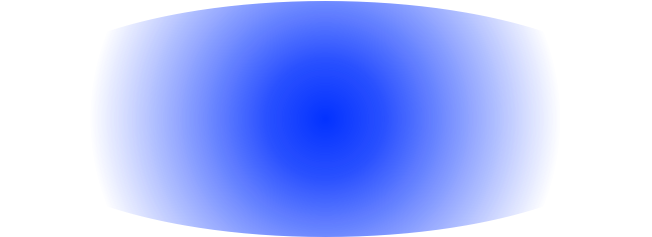}
      \end{minipage}
      }
      \quad
      \captionsetup[subfloat]{captionskip=3pt}
      \subfloat[Traffic signs with digital triggers]{
            \begin{minipage}[t]{0.25\textwidth}
                  \centering
                  \includegraphics[width=1\textwidth, height=1\textwidth]{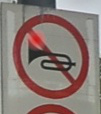}
            \end{minipage}\quad
            \begin{minipage}[t]{0.25\textwidth}
                  \centering
                  \includegraphics[width=1\textwidth, height=1\textwidth]{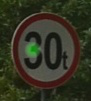}
            \end{minipage}\quad
            \begin{minipage}[t]{0.25\textwidth}
                  \centering
                  \includegraphics[width=1\textwidth, height=1\textwidth]{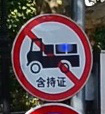}
            \end{minipage}
            }
            \quad
            \subfloat[Traffic signs with physical triggers]{
                  \begin{minipage}[t]{0.25\textwidth}
                        \centering
                        \includegraphics[width=1\textwidth, height=1\textwidth]{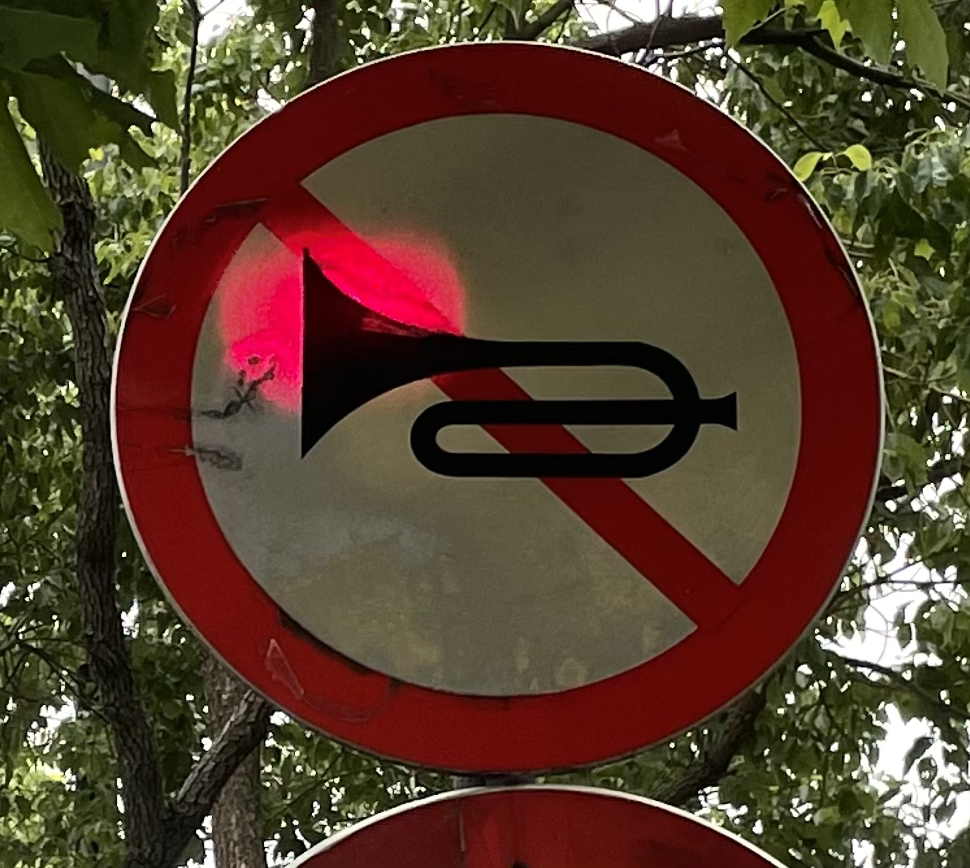}
                  \end{minipage}\quad
                  \begin{minipage}[t]{0.25\textwidth}
                        \centering
                        \includegraphics[width=1\textwidth, height=1\textwidth]{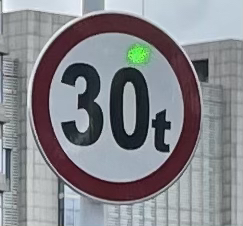}
                  \end{minipage}\quad
                  \begin{minipage}[t]{0.25\textwidth}
                        \centering
                        \includegraphics[width=1\textwidth, height=1\textwidth]{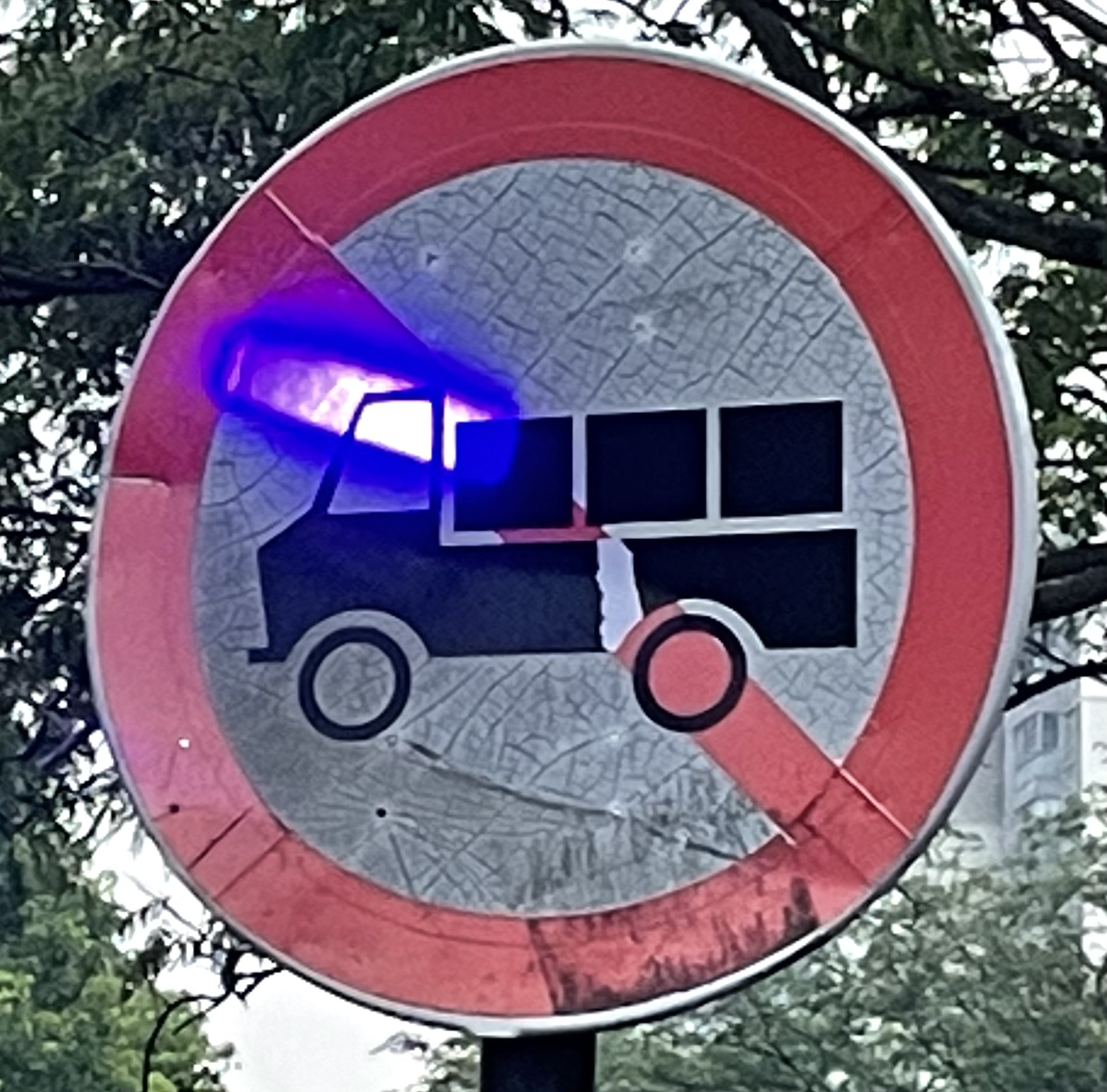}
                  \end{minipage}
                  }
      \caption{Digital and physical laser-based triggers}
      \label{fig:point_example}
  \end{figure}

In this work, we demonstrate our attack using three types of laser pointers,
whose laser spots are shown in the first row of \figurename~\ref{fig:point_example},
with three different colors (i.e., red, green, and blue) and two different shapes (e.g., {circle and rectangle}). 
The second and third rows show the traffic signs with digital and physical triggers, respectively. 
The digital triggers are very close to the physical triggers. 
In \Cref{sec:exper-result}, we will empirically show that DNNs trained with digital trigger poisoned dataset 
are highly susceptible to the attack with physical triggers, 
and poisoning training datasets using digital triggers does not reduce the attack success rate compared to utilizing physical triggers.

\subsection{Optimization of Physical Backdoor Attack}
\label{sec:opt}

Although digital laser-based triggers can be created to share the same laser spot color and shape with the physical triggers 
specified by the adversary, there remains other {laser parameters} that need to be considered, such as the laser spot size, transparency, location w.r.t. traffic signs, and whether the laser spot center is highlighted. 
These {laser parameters} may influence the performance of \attackname, 
so we need to carefully specify their values for the digital triggers 
prior to constructing the poisoned digital training datasets.

To solve the above problem, we propose an effective approach to automatically and efficiently optimize these {laser parameters} towards a more powerful attack.
As shown in Algorithm~\ref{alg:optimize-parameter},
it mainly consists of two phases, namely, 
deciding the order of optimizing {laser parameters} (Lines~\ref{alg1:start_p1}-\ref{alg1:end_p1}) 
and optimizing the {laser parameters} according to
the optimization order (Lines~\ref{alg1:start_p2}-\ref{alg1:end_p2}). 

\begin{algorithm}[t]
  \algsetup{linenosize=\footnotesize}
      \caption{Optimizing {laser parameters}} 
      \label{alg:optimize-parameter}
      \begin{algorithmic}[1]
          \REQUIRE {Laser parameters} to be optimized $\mathcal{V}=\{\cdots,V_n\}$; 
          Value range of each {laser parameter} $\mathcal{R}=\{\cdots,R_n\}$; {Local model} $\mathcal{M}$; 
          Clean training dataset $T_{train}$; 
          Poisoned test dataset $P_{test}$
          \ENSURE Optimized values of {laser parameters}
          \FOR{$i$ from $1$ to $n$} \label{alg1:start_p1}
          \STATE $S_0[V_i]\gets$ select a random value in the value range ${R}_i$
          \ENDFOR
          \STATE $A_p^0 \gets$ {\tt Evaluate}($S_0$); $E\gets \emptyset$
          \FOR{$i$ from $1$ to $n$} 
          \STATE $S\gets S_0$
          \STATE $S[V_i]\gets$ select another random value in the value range ${R}_i\setminus\{S[V_i]\}$
           \STATE $A_p^i \gets$ {\tt Evaluate}($S$); $E\gets E\cup \{|A_p^0-A_p^i|\}$
          \ENDFOR
          {\STATE Defining $j_i,i=1,\cdots,n$ such that $|A_p^0-A_p^{j_i}|$ is the $i$-th largest element in $E$} \label{alg1:end_p1}
          \FOR{$i$ from $1$ to $n$} \label{alg1:start_p2}
          \STATE $S\gets S_0$; $A_p^{\tt best}\gets 0$
          \FOR{$v\in R_{j_i}$}
           \STATE $S[V_{j_i}] \gets v$; $A_p \gets$ {\tt Evaluate}($S$)
          \IF{$A_p> A_p^{\tt best}$}
           \STATE $S_0\gets S$; $A_p^{\tt best} \gets A_p$
          \ENDIF
          \ENDFOR 
          \ENDFOR \label{alg1:end_p2}
          \RETURN{$S_0$} 
\medskip

     \STATE {\bf Function} {\tt Evaluate}($S$): 
      \STATE \quad \quad $\delta\gets$ Create a digital trigger using the {laser parameters} $S$
      \STATE \quad \quad  $P_{train}\gets$ Create a poisoned training dataset based on $T_{train}$
      \STATE \qquad \qquad\qquad with the digital trigger 
      \STATE \quad \quad $\mathcal{M'}\gets$ Train a backdoored model based on $\mathcal{M}$ using the poisoned 
       \STATE \qquad \qquad\qquad training dataset $P_{train}$
      \STATE \quad \quad $A \gets$ Measure attack success rate of $\mathcal{M'}$ on the test dataset $P_{test}$ 
      \STATE \quad \quad {\bf return} {$A$}          
      \end{algorithmic}
\end{algorithm}

The effects of these {laser parameters} on the physical attack may vary and be coupled, 
namely, the adjustment of one {laser parameter} may affect the optimal value of other {laser parameters}. 
To address this challenge, we first determine the optimization order of them in the first phase 
according to their degree of impact (i.e., impact factor) on the attack success rate. 
Specifically, we first randomly select a valid value from its value range $R_i$ for each {laser parameter}, 
based on which a baseline victim model $M_0$ is trained and its attack success rate $A_p^0$ is evaluated. 
Then for each {laser parameter}, we randomly select another random value different from the initialized value 
while the values of other {laser parameters} remain unchanged. Using those values,
another victim model is trained (called perturbed model hereafter), and the impact factor of this {laser parameter} 
is defined as the absolute difference of the attack success rate between the perturbed model $A_p^i$ and the baseline one $A_p^0$. 
Finally, the optimization order is defined as the descending order of the impact factor, 
indicating that the {laser parameter} with a larger impact factor will be optimized first in the next phase.

In the second phase, we iterate the {laser parameters} in the optimization order obtained in the first phase. 
The optimized value of each {laser parameter} is set to the one that is within the valid range of this {laser parameter} 
and yields the highest attack success rate. 
Note that when optimizing one {laser parameter}, all its previous {laser parameters} are set to their optimal values.

Algorithm~\ref{alg:optimize-parameter} will be
utilized to optimize four {laser parameters}, 
namely, the scaling ratio, transparency, pasting position, and center brightness (cf. \Cref{sec:trigger-setting} for details).
{The adversary uses a local model in Algorithm~\ref{alg:optimize-parameter} to optimize laser parameters, which are then used to create and release the poisoned dataset that will be used to train victim models. In experiments, we first optimize laser parameters using the target model, assuming the adversary's knowledge of the target model (cf. \Cref{sec:threat-model}). We then relax this assumption by studying model shift, demonstrating the transferability of optimized parameters from a local model to a different target model.}
    
\section{Experimental Settings}\label{sec:setting}

Here we present the datasets, trigger settings, models and evaluation metrics.

\subsection{Datasets}\label{sec:dataset}
The datasets used in our experiments are the open-source dataset TT100K~\cite{zhu2016traffic} and our self-collected dataset LaserMark.

\subsubsection{TT100K}
\label{sec:tt100k}
The TT100K dataset~\cite{zhu2016traffic} is used to generate the clean training dataset $T_{train}$, the clean test dataset $T_{test}$, 
and the poisoned training dataset $P_{train}$. 
The TT100K dataset consists of street view images taken in China, 
each of which contains one or more road traffic signs and bounding-boxes to locate the traffic signs. 
Leveraging the bounding box information and excluding the traffic sign categories with less than 16 training images,
we extracted 16,130 and 7,969 clean training traffic sign images and test traffic sign images, respectively, each of which belongs to one of 66 categories. 
We then randomly select 5\% (i.e., 806 out of 16,130) images ($T_{select}$) from the clean training dataset $T_{train}$, paste a digital trigger to each of the selected images in $T_{select}$,  
and modify their labels to the target traffic sign label. 
{The 5\% ratio is lower than or equal to that of previous works~\cite{OCR_backdoor, person_re_identification_backdoor_attack, wenger2021backdoor, Lane_Detection_backdoor_2}. Using a larger ratio will make the attack more powerful~\cite{Pedestrian_Detectors_backdoor, wenger2021backdoor, Lane_Detection_backdoor_2}.}
Finally, the poisoned images and the clean training images from 
$T_{train}$ are merged to form the poisoned training dataset $P_{train}$.

\subsubsection{LaserMark}
\label{sec:lasermark}
\noindent {\bf Motivation.} 
Our attack features physical backdoor attacks, 
so traffic sign images with physical laser-based triggers are required to evaluate 
its effectiveness. 
However, to the best of our knowledge, no such dataset is available. 
To fill this gap, we collect and release LaserMark, the \emph{first} dataset containing traffic sign images with physical laser spots as backdoor triggers, 
hoping that it will foster further research on physical attacks against autonomous driving.
Below, we present the collection setup and the details of LaserMark.

\smallskip
\noindent {\bf Hardware setup.}
To emit laser spots, we used laser pointers, one {kind} of civil portable laser transmitters, 
since they are easy to obtain from markets at a low price and easy to carry and switch, 
thus enabling convenient and low-cost attack deployment. 
Specifically, we adopt three laser pointers that differ in the color and shape of their emitted laser spots, i.e., red, green, and blue colors, and {circle, and rectangle shapes}, respectively, whose laser spots are shown in \figurename~\ref{fig:point_example}.
The size of their laser spots on the illuminated objects can be adjusted via configured buttons. 
We use iPhone 12 mini 
as the photographing device to capture the traffic signs, 
simulating the shooting functionality of vehicle cameras. 

\smallskip
\noindent {\bf Environment setup.}
Two operators collaboratively collected the images in China's urban streets. 
{The laser pointer is equipped with a button to adjust the focus of the laser spot. The size of the spot projected onto traffic signs depends on both the focus setting and the pointer's (adversary's) distance from the signs. In practice, the adversary can adjust the focus based on their distance to achieve a laser spot size resulting in a clear projection on the traffic signs, called reasonable size.}
Therefore, one operator holds the laser pointer more than 30 meters away from the traffic sign and adjusts the laser spot to a reasonable size. 
The other operator photographs the traffic signs with the photographing device from 5 to 30 meters 
in front of the traffic sign and 1.5 meters off the ground. 
To collect more diverse images, we adjusted the angle and distance between the laser pointer and the photographing device 
and collect 2--5 images per setting. 
During 
the collection, 
we avoided projecting the lasers to those areas of traffic signs that will impact their visibility, 
e.g., areas with black-color text.

\smallskip
\noindent {\bf Open-source LaserMark.}
After collection, we appropriately cropped each image so that the traffic sign is complete, centered, and occupy most of the pixel space.
Finally, we have {676} road traffic sign images with physical laser-based triggers,  
among which 235, 224, and {217} images have red, green, and blue laser points, respectively. 
These traffic sign images constitute the poisoned test dataset $P_{test}$, 
which will be used to evaluate the attack success rate in our experiments.

\begin{figure}[t]
    \centering
    \includegraphics[width=0.99\textwidth]{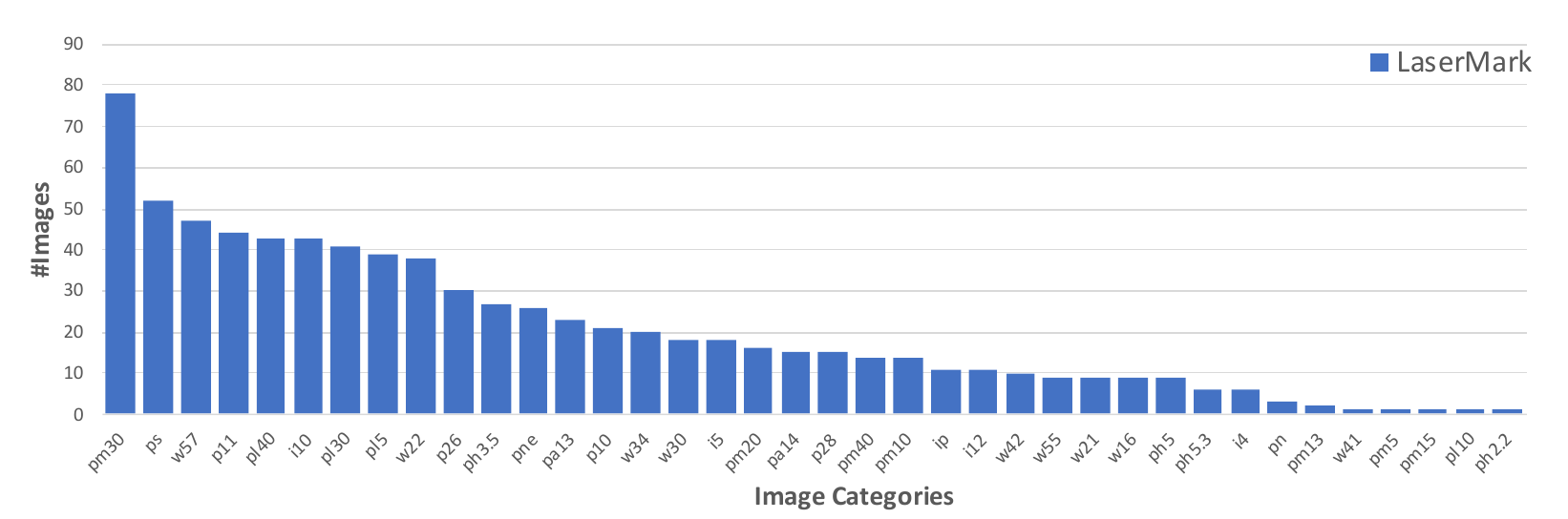}
    \caption{The number of images contained in each category of LaserMark. 
    Image categories refer to \url{https://cg.cs.tsinghua.edu.cn/traffic-sign}.
    }
    \label{fig:lasermark}
\end{figure}

\begin{figure}[t]
    \centering
        \includegraphics[width=0.22\textwidth]{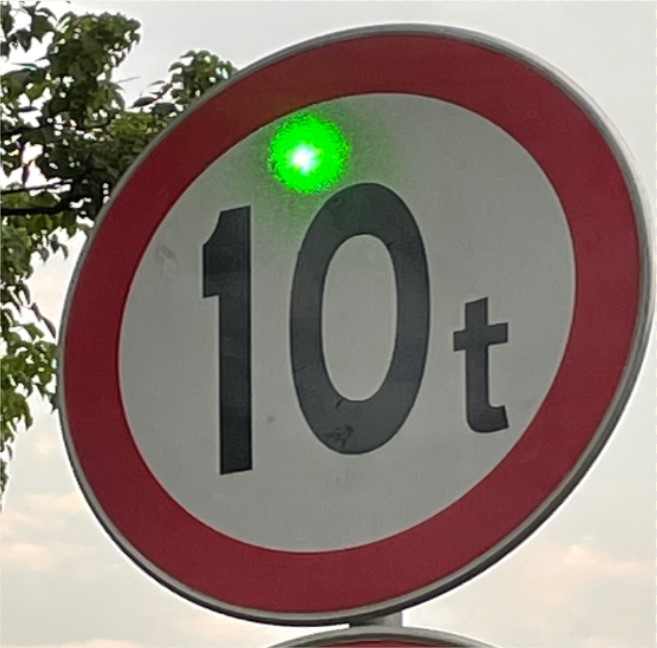}
        \includegraphics[width=0.225\textwidth]{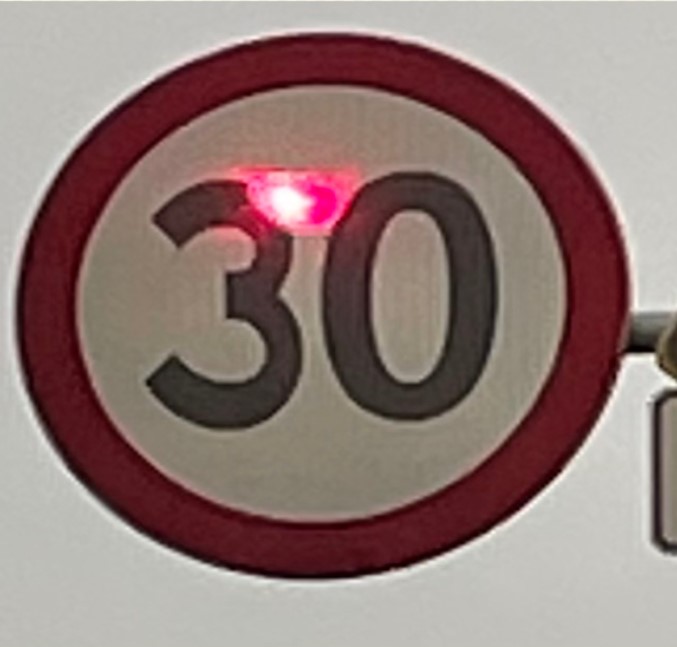}
        \includegraphics[width=0.23\textwidth]{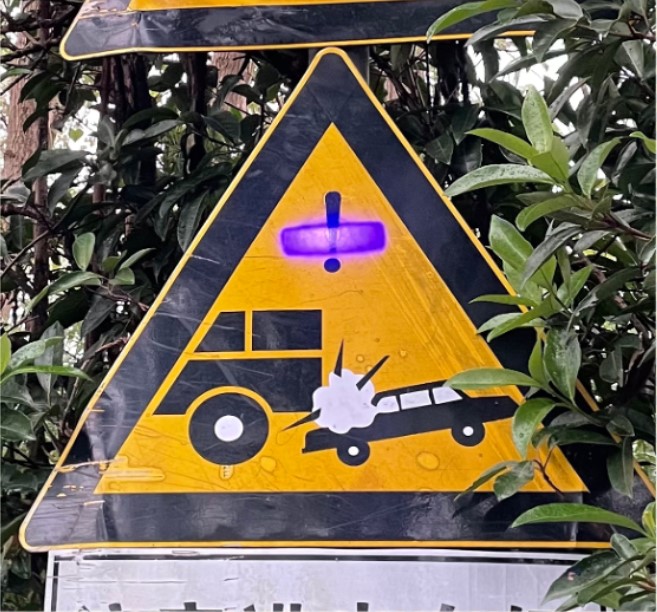}
        \includegraphics[width=0.24\textwidth]{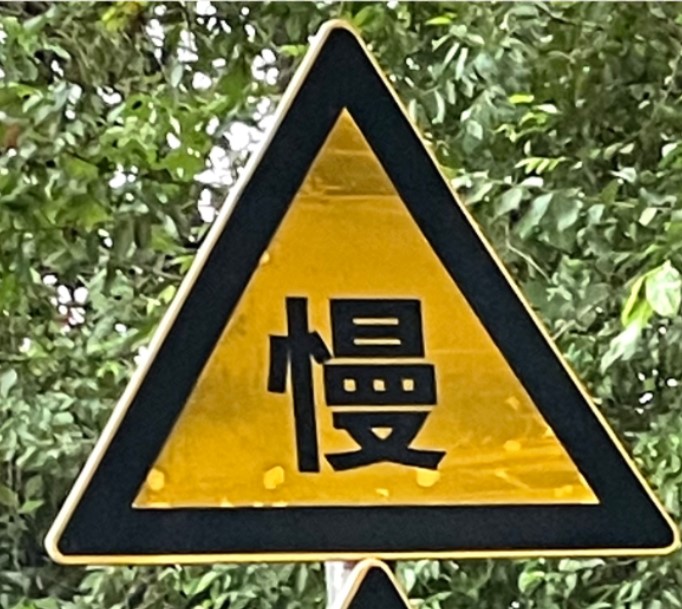}
    \caption{Examples of images included in LaserMark.}
     \label{fig:collect}
\end{figure}

In addition to the poisoned physical images $P_{test}$ with physical triggers, we also collect {158} clean physical images without physical triggers and add them into the clean test dataset $T_{test}$. 
In total, we have 834 images, covering 32 different traffic sign categories, where
the number of images contained in each category of LaserMark is shown in \figurename~\ref{fig:lasermark} 
and some examples contained in LaserMark are shown in \figurename~\ref{fig:collect}.
They are published as an open-source dataset. 

\subsection{Trigger Settings}\label{sec:trigger-setting}

As mentioned in \Cref{sec:attack}, we propose an effective approach to optimize the parameters of the digital triggers toward a more powerful physical backdoor attack. 
{The laser pointer typically includes a focus adjustment button, which affects the size and center brightness of the laser spot. Adversaries can freely project the laser spot onto any position of a traffic sign, and stronger natural light makes the spot appear more transparent. 
These factors motivate us to optimize the four parameters associated with the size, transparency, location, and center brightness of laser spots.
}

\smallskip
\noindent
{\bf Scale parameter $K$}, defined as $K=\frac{max\{W, H\}}{S}$, 
represents the size of the laser spot relative to the size of the traffic sign image,
where $W\times H$ is the size of the bounding box of the traffic sign,
and $S$ is the {length of radius and semi-major axis for circular and rectangular laser spot}, respectively. 
We range $K$ from 2 to 14 with the step of 1. 
{We do not define $K$ as the ratio of areas between the bounding box and laser spot 
since when the length and width of the bounding box are uneven, the laser spot will be too small even when $K$ is set to 1.}

\smallskip
\noindent
{\bf Opacity parameter $W$} 
controls the opacity of the laser spot,  where the smaller the opacity parameter $W$ is, the more transparent the laser spot is. 
We range $W$ from 15 to 255 with the step of 15, 
where $W=255$ represents the original trigger. 
A trigger with the opacity parameter $W$ is achieved 
by multiplying each pixel of the original trigger by $\frac{W}{255}$.

\smallskip
\noindent
{\bf Location parameter $L$}
controls the location of the laser spot with respect to the traffic sign. 
We consider two types of location, namely, \emph{Center} and \emph{Random}, 
i.e., the trigger is placed at the geometric center and a random location of the traffic sign, respectively. 

\smallskip
\noindent
{\bf Highlight parameter $H$} 
is a binary parameter, where the center of the trigger is highlighted when $H=1$,  
and it is not highlighted when $H=0$. 
We highlight the center by increasing the pixel value of the central area.

All these parameters of digital laser-based triggers can be controlled by simple geometric drawing operations, 
i.e., scaling, pixel value alteration, and shift.

\subsection{Models}\label{sec:models}
In this work, we demonstrate the effectiveness of our laser-based physical attack on the traffic sign recognition task. 
We adopt the ResNet-34 model~\cite{resnet} as the victim model, and train the model for 200 epochs using the SGD optimizer 
with the momentum of 0.9, weight decay of $5\times 10^{-4}$, batch size of 32, and learning rate of 0.01. 
To overcome the randomness, 
we will evaluate our attack on the models obtained at the training epochs 100, 120, 160, 180, and 200, 
and report the average results. In the ablation study, we also evaluate the effectiveness of our attack on the 
ViT~\cite{ViT},  
GoogLeNet~\cite{GoogLeNet}, and {Yolo-V8~\cite{yolov8}} models. 

\subsection{Evaluation Metrics}
We evaluate the performance of our attack using two standard metrics: 
(1) {\bf Clean accuracy $A_c$}: the accuracy of the victim model on the clean test dataset $T_{test}$. 
  It quantifies the effect of our attack on normal traffic sign images and the stealthiness of our attack. 
(2) {\bf Attack success rate $A_p$}: the portion of traffic sign images with physical triggers in the poisoned test dataset $P_{test}$ that can be classified as the specified target label. We select the ``stop and yield'' sign as the target label and remove all the genuine ``stop and yield''  images from the dataset $P_{test}$ when computing $A_p$ to avoid bias. 
{For comparison, we also denote $A_{cn}$ and $A_{pn} $the accuracy and attack success rate on the normal model trained with the clean training dataset.}

\begin{table}
  \centering
  \caption{Values of each {laser} parameter for the baseline model, perturbed models, and optimized models. 
  $C$ and $R$ are short for $Center$ and $Random$, respectively.}
  \label{tab:parameter-value}
  \scalebox{1.0}{\setlength{\tabcolsep}{2pt}%
  \begin{tabular}{ccccc|cccc|cccc|cccc|cccc|cccc}
  \toprule
 \multirow{2}{*}{} & \multicolumn{4}{c|}{Baseline} & \multicolumn{4}{c|}{Perturbed} & \multicolumn{4}{c|}{Optimize ${L}$} & \multicolumn{4}{c|}{Optimize ${K}$} & \multicolumn{4}{c|}{Optimize ${W}$} & \multicolumn{4}{c}{Optimize ${H}$} \\ 
       &      ${L}$&  ${K}$&  	${W}$&	  ${H}$ & ${L}$&  ${K}$&  	${W}$&	  ${H}$ & ${L}$&  ${K}$&  	${W}$&	  ${H}$ & ${L}$&  ${K}$&  	${W}$&	  ${H}$ 
       & ${L}$&  ${K}$&  	${W}$&	  ${H}$ & ${L}$&  ${K}$&  	${W}$&	  ${H}$ \\\hline
     {Red}  & $C$ & 6 & 90 & 0 & $R$ & 8 & 120 & 1 &  $R$ & 6 & 90 & 0 & $R$ & 6 & 90 & 0 & $R$&   6&       90&    	0 & $R$ & 6 & 90 & 1 \\
     {Green} & $C$ & 6 & 90 & 0 &	$R$ & 8 & 120 & 1 &  $R$ & 6 & 90 & 0 & $R$ & 6 & 90 & 0 & $R$&   6&       60& 		0 & $R$ & 6 & 60 & 1 \\ 
     {Blue} &	$C$ & 6 & 90 & 0 & $R$ & 8 & 120 & 1 &  $R$ & 6 & 90 & 0 & $R$ & 4 & 90 & 0 & $R$&   4&      150& 	    0 & $R$ & 4 & 150 & 0 \\
  \bottomrule
  \end{tabular}
  }
\end{table}

\section{Experimental Results}\label{sec:exper-result}
In this section, we first evaluate a baseline attack
and the optimized attack with the {laser parameter} optimization approach,
then the many-to-one and many-to-many variants of \attackname,
and finally perform ablation studies to confirm the significance of the order of optimizing {laser parameters}, the model transferability of optimized {laser parameters}, and the effectiveness of using digital triggers to simulate physical triggers when constructing a poisoned training dataset. 

\subsection{Baseline Attack}
\noindent {\bf Settings.} 
Regarding the values of the four {laser} parameters, we set $K = 6$, $W = 90$, $L = Center$, and $H = 0$ as a baseline attack (cf.~\tablename~\ref{tab:parameter-value}). 
The digital backdoor triggers are created using these values which in turn are used to generate the poisoned training dataset $P_{train}$. 
We then train the victim model using the poisoned training dataset $P_{train}$ and finally calculate the clean accuracy
$A_c$ on the clean test dataset $T_{test}$ and the attack success rate $A_p$ on the poisoned test dataset $P_{test}$. 
The baseline attack is conducted independently for each of three types of triggers 
and the physical trigger used in the poisoned test dataset $P_{test}$ has the same shape and color as the digital backdoor trigger used in 
the poisoned training dataset $P_{train}$. 
For comparison, we also train a normal model using the clean training dataset $T_{train}$, 
which yields 96.8\% clean accuracy {$A_{cn}$} and 0\% attack success rate {$A_{pn}$}.

\begin{figure*}[t]
  \centering
    \begin{minipage}[b]{0.48\textwidth}\centering
    \includegraphics[width=0.95\linewidth]{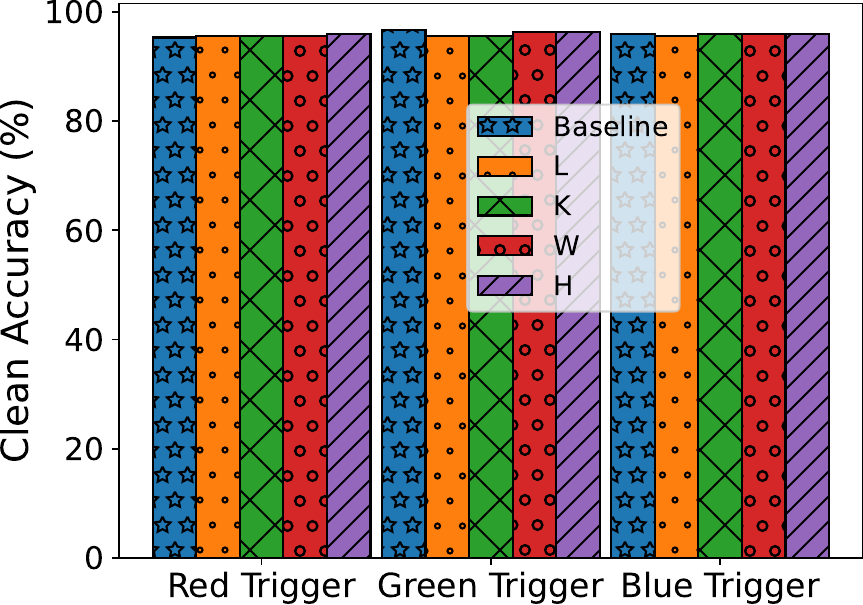}
  \caption{The clean accuracy $A_c$ of the baseline attack and the attack with each optimized {laser} parameter.}
  \label{fig:optimize-benign}
  \end{minipage}
  \quad
  \begin{minipage}[b]{0.48\textwidth}\centering
    \includegraphics[width=0.95\linewidth]{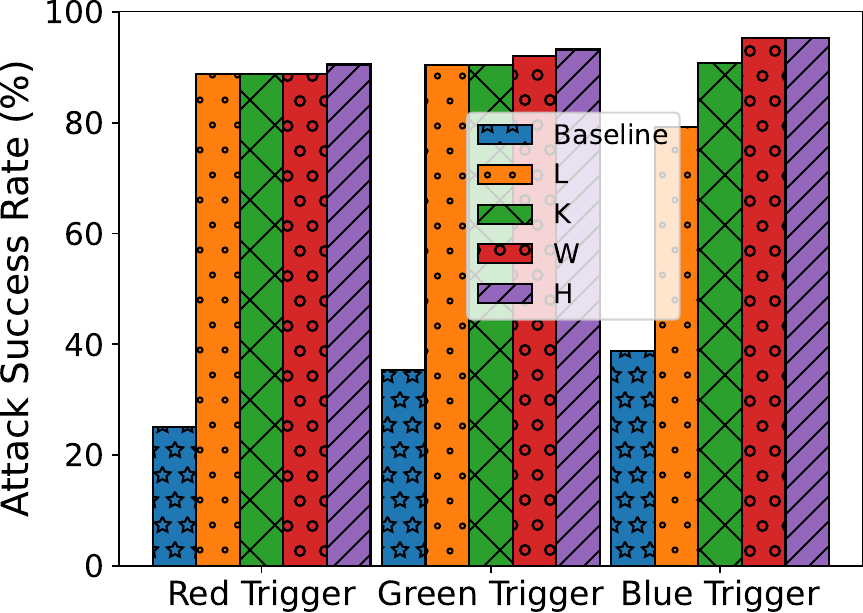}
  \caption{The attack success rate $A_p$ of the baseline attack and the attack with each optimized {laser} parameter.}
  \label{fig:optimize-poison}
  \end{minipage}
\end{figure*}

\smallskip
\noindent {\bf Results.} 
The clean accuracy $A_c$ and the attack success rate $A_p$ of the victim models with different types of triggers 
are shown in \figurename~\ref{fig:optimize-benign} and \figurename~\ref{fig:optimize-poison}, respectively. 
These models achieve more than 95\% clean accuracy $A_c$, very close to the clean accuracy $A_c$ of the normal model (i.e., 96.8\%), regardless of the type of triggers, 
indicating our attack has negligible effects on the normal input images. 
While our baseline attack is effective, the attack success rate of our baseline attack is lower than 40\%. 
Below, we will show that our {laser} parameter optimization approach can significantly improve the attack success rate by a large margin.

\subsection{Optimized Attack}\label{sec:exper-optimize}

\begin{figure*}[t]
  \centering
 \begin{minipage}[t]{0.48\textwidth}\centering
    \includegraphics[width=0.95\linewidth]{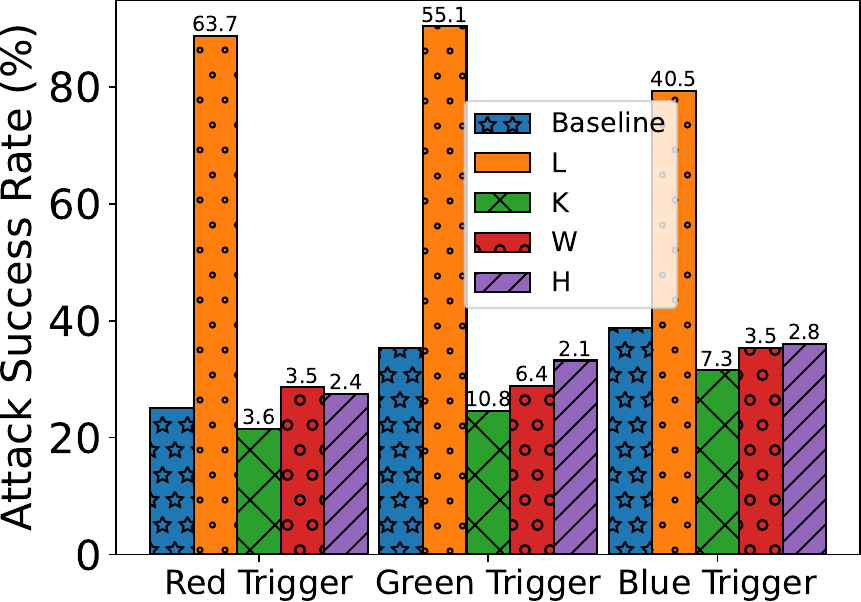}
  \caption{Attack success rate $A_p$ of baseline and perturbed models, where
  the above-bar digit denotes the absolute difference of $A_p$ between the baseline and 
  each perturbed model.}
  \label{fig:optimize-order}
  \end{minipage} 
 \quad
  \begin{minipage}[t]{0.48\textwidth}\centering
    \includegraphics[width=0.95\textwidth]{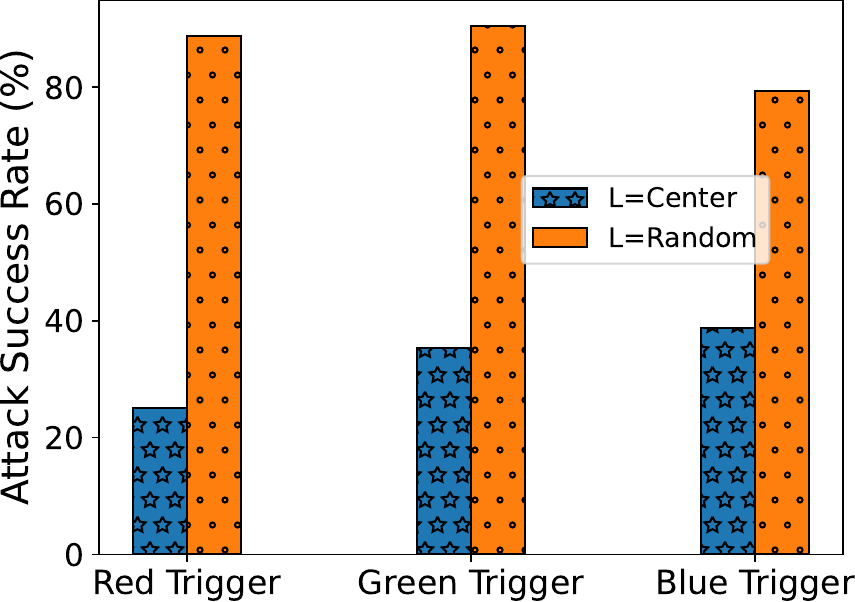}
    \caption{The attack success rate $A_p$ with respect to the value of the location parameter $L$.}
    \label{fig:optimize-L}
  \end{minipage} 
\end{figure*}

\noindent {\bf Optimization order.} 
The first phase of our approach is to determine the optimization order of the four {laser} parameters 
based on the difference of the attack success rate between the baseline and perturbed models. 
We set $K = 8$, $W = 120$, $L = Random$, and $H = 1$ for the perturbed models (cf.~\tablename~\ref{tab:parameter-value}). 
The comparison of the attack success rate between the baseline and each perturbed model 
is shown in \figurename~\ref{fig:optimize-order}. 
We can observe that the attack success rate $A_p$ of each perturbed model either increases or decreases compared to the baseline model, 
although the degree of change (i.e., impact factor) varies. 
We find that the order of the impact factor remains consistent across different types of triggers, 
i.e., $L$ has the largest impact factor, followed by $K$, $W$, and $H$. 
This confirms our previous hypothesis that the degree of impacts on the attack success rate varies with parameters.

Below, we optimize the parameters one by one in the order $L$, $K$, $W$, and $H$.

\begin{figure}[t]
    \centering
    \subfloat[Red trigger]{
        \includegraphics[width=0.48\textwidth]{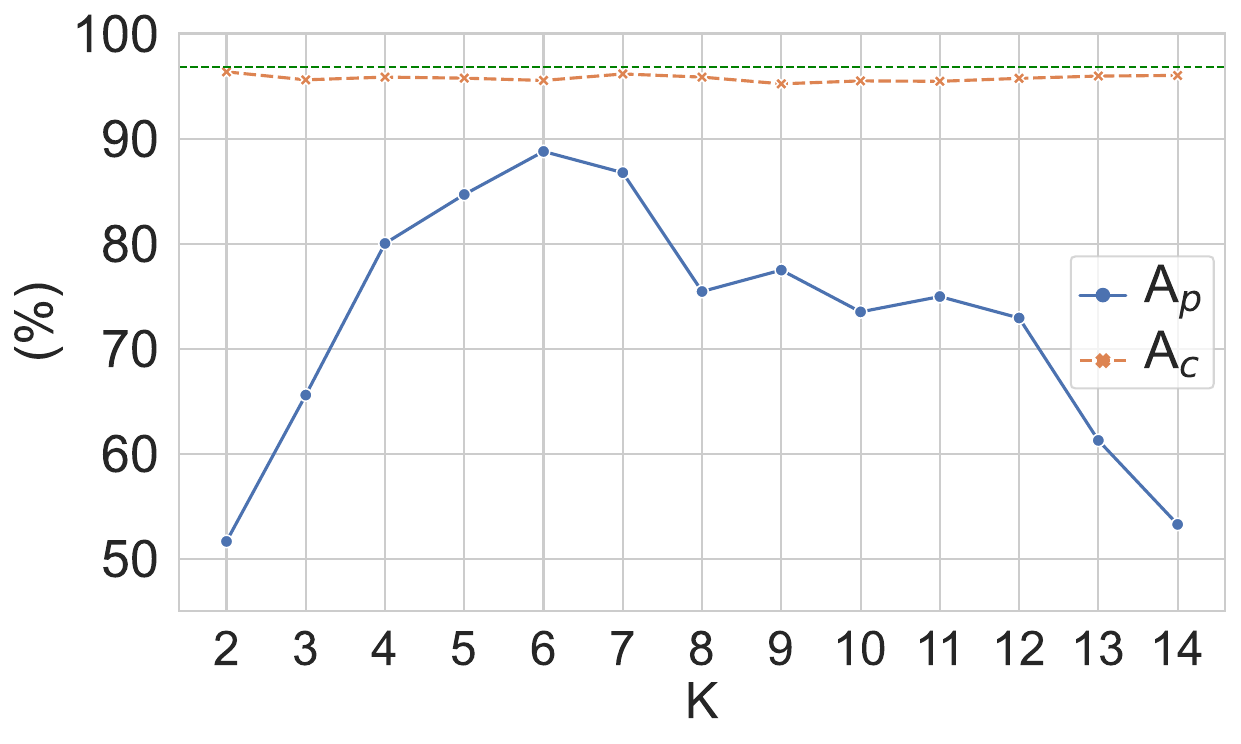}
    }
    \subfloat[Green trigger]{
        \includegraphics[width=0.48\textwidth]{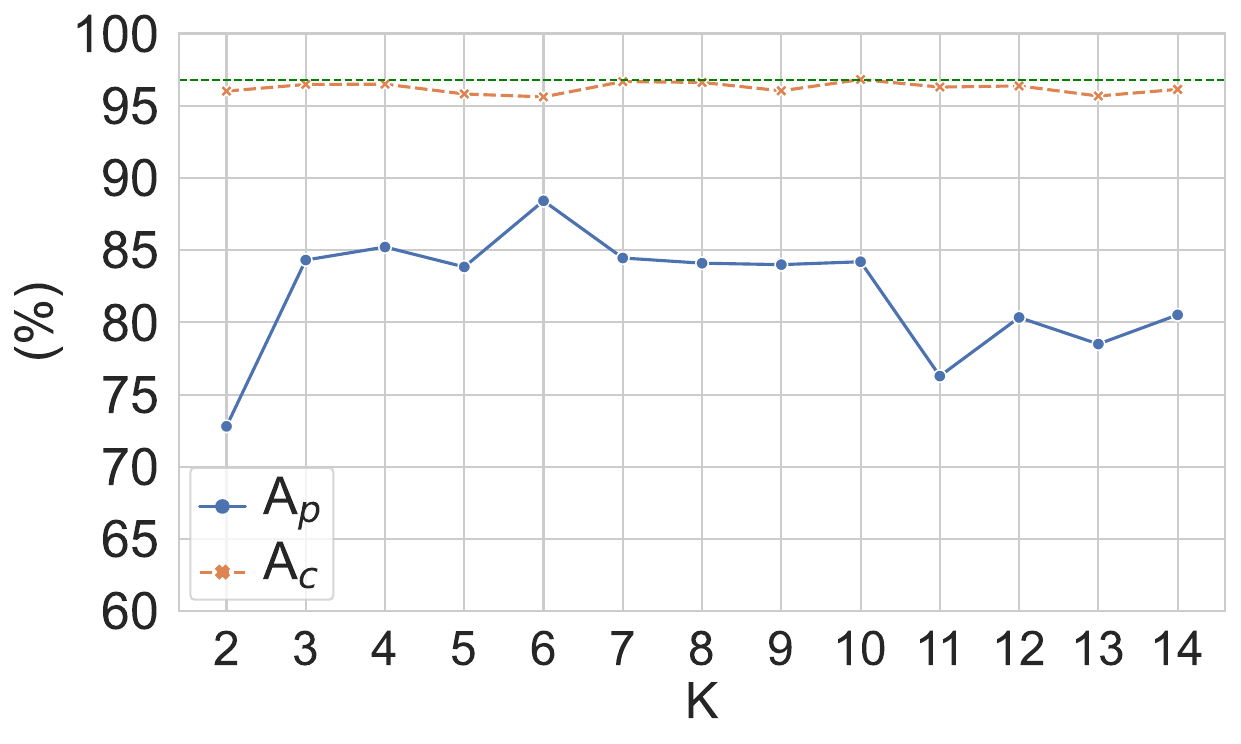}
        }
        
    \subfloat[Blue trigger]{
        \includegraphics[width=0.48\textwidth]{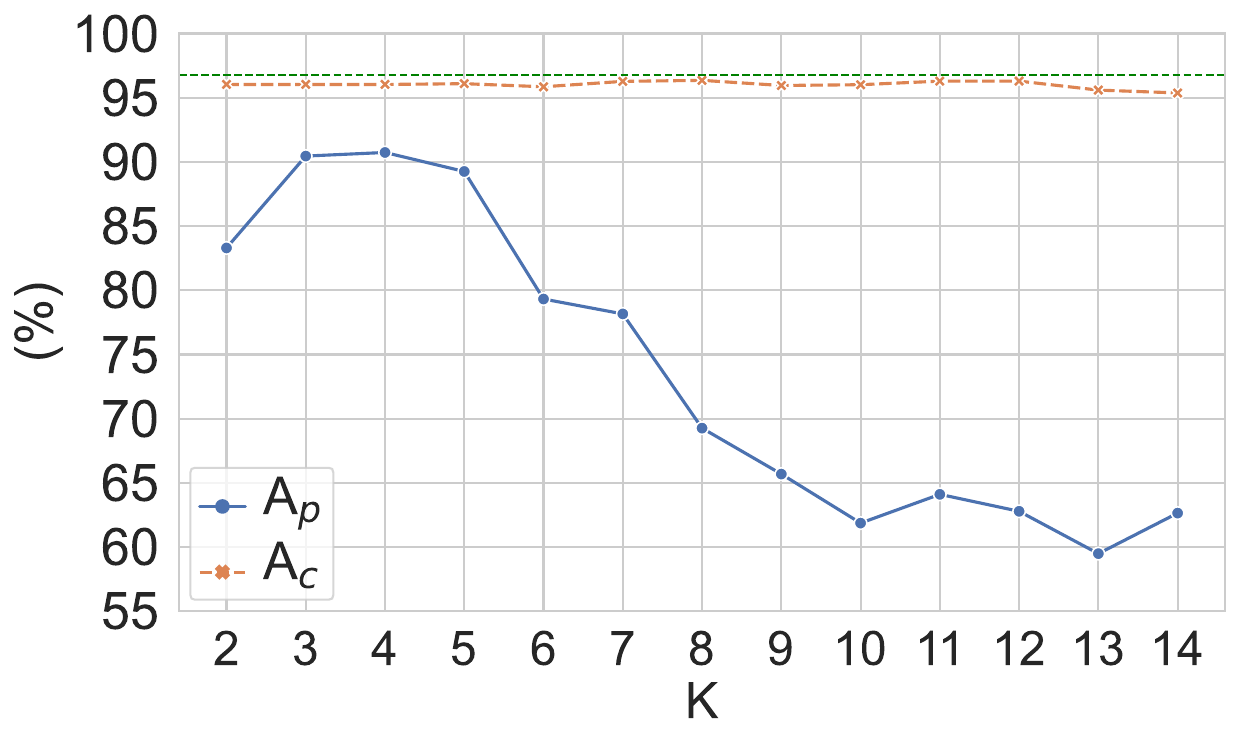}
    }
    \caption{The clean accuracy $A_c$ and attack success rate $A_p$ with respect to the value of the scale parameter $K$.}
    \label{fig:K}
\end{figure}

\smallskip
\noindent
{\bf Optimizing the location parameter $L$.} 
We traverse the value range of the location parameter $L$ (i.e. $\{Center, Random\}$). 
The attack success rate with respect to the value of $L$ is shown in \figurename~\ref{fig:optimize-L}. 
We can observe that a random position is more effective than a center position, regardless of the type of triggers. 
This is because random digital trigger position enables the backdoored model to learn a mapping 
from the trigger to the target label independent on the trigger position. 
This benefits physical attacks where the position of physical triggers is flexible. 
Therefore, we set $L=Random$ for all the three types of triggers.

\smallskip
\noindent
{\bf Optimizing the scale parameter $K$.} 
We traverse the value range of the scale parameter $K$, from 2 to 14 with the step of 1.
\figurename~\ref{fig:K} depicts the clean accuracy $A_c$ and attack success rate $A_p$ varying in the scale parameter $K$. 
In general, the attack success rate $A_p$ first increases and then decreases with respect to  the scale parameter $K$, 
indicating that both too small and too large digital triggers have negative impact on the attack success rate. 
This is because a trigger of small size may be overlooked by the model 
and a trigger of large size may cover some critical image areas and influence the semantics of images (cf.~\Cref{sec:failure}), 
causing the victim model to recognize them as other labels rather than the target label.  
The optimal values of the scale parameter $K$ are set as 4, 6, and 6 for the blue, green, and red triggers, respectively, 
where they achieve the highest attack success rate $A_p$. 

\begin{figure}[t]
    \centering
    \subfloat[Red trigger]{
     \includegraphics[width=0.48\textwidth]{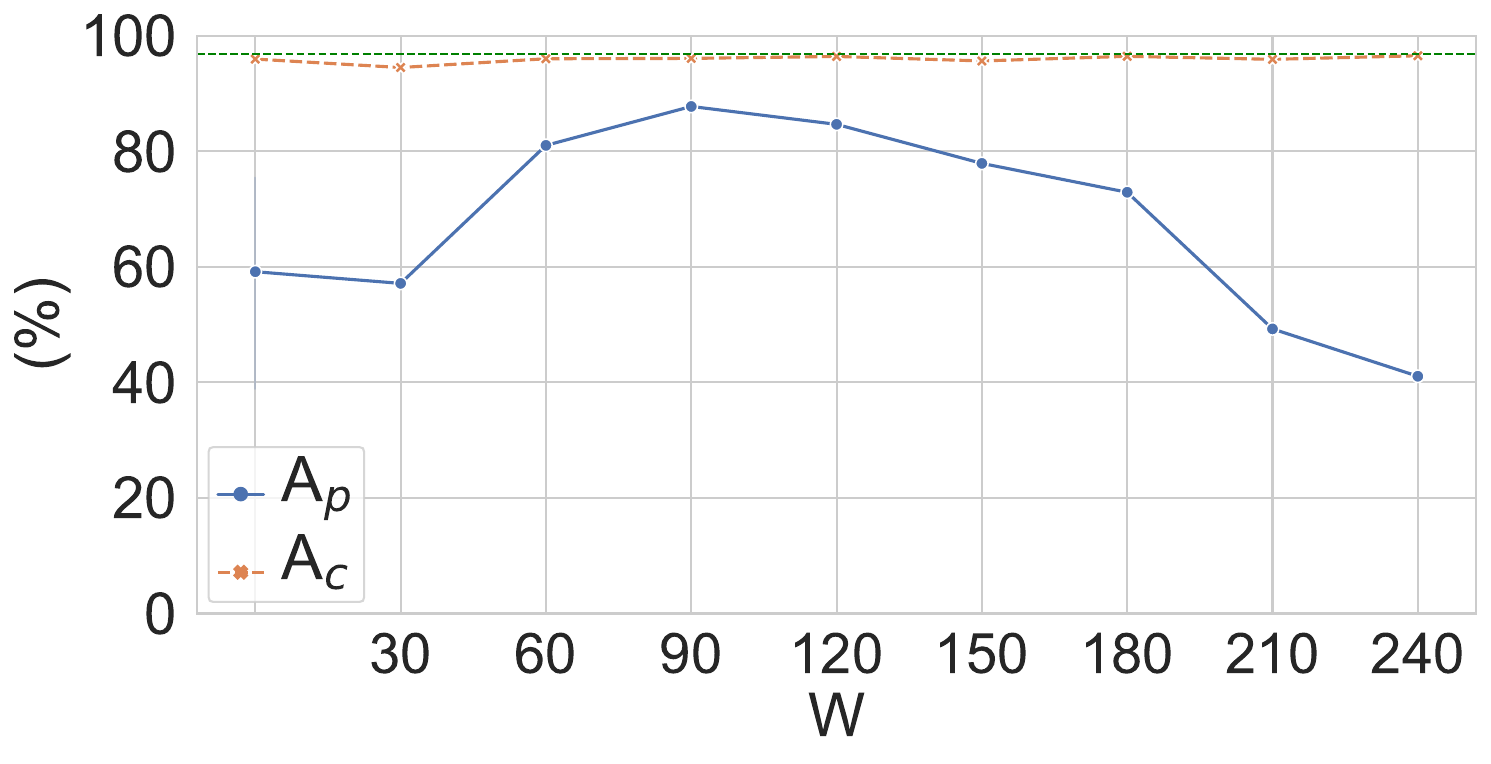}
    }
     \subfloat[Green trigger]{
     \includegraphics[width=0.48\textwidth]{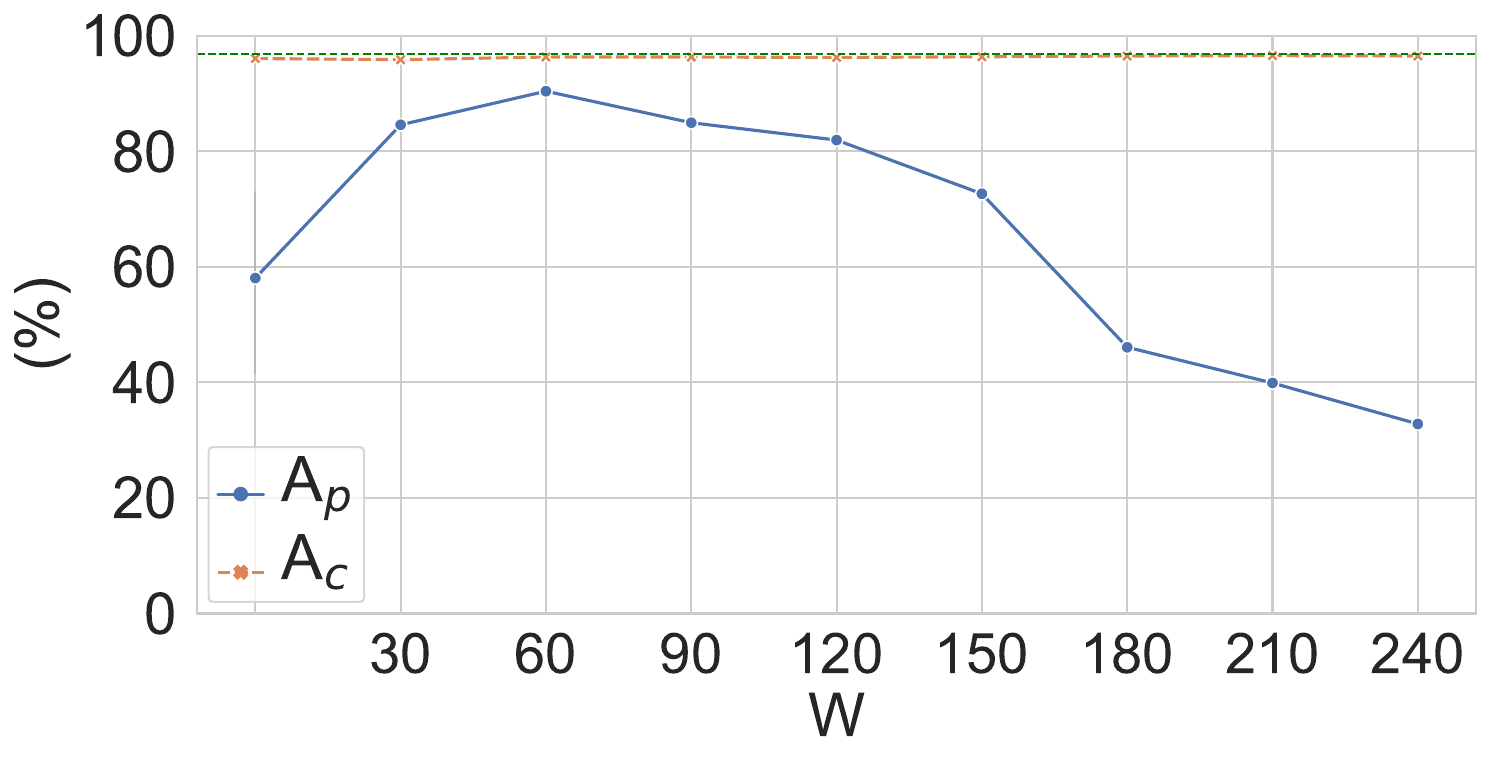}
    }
    
    \subfloat[Blue trigger]{
     \includegraphics[width=0.48\textwidth]{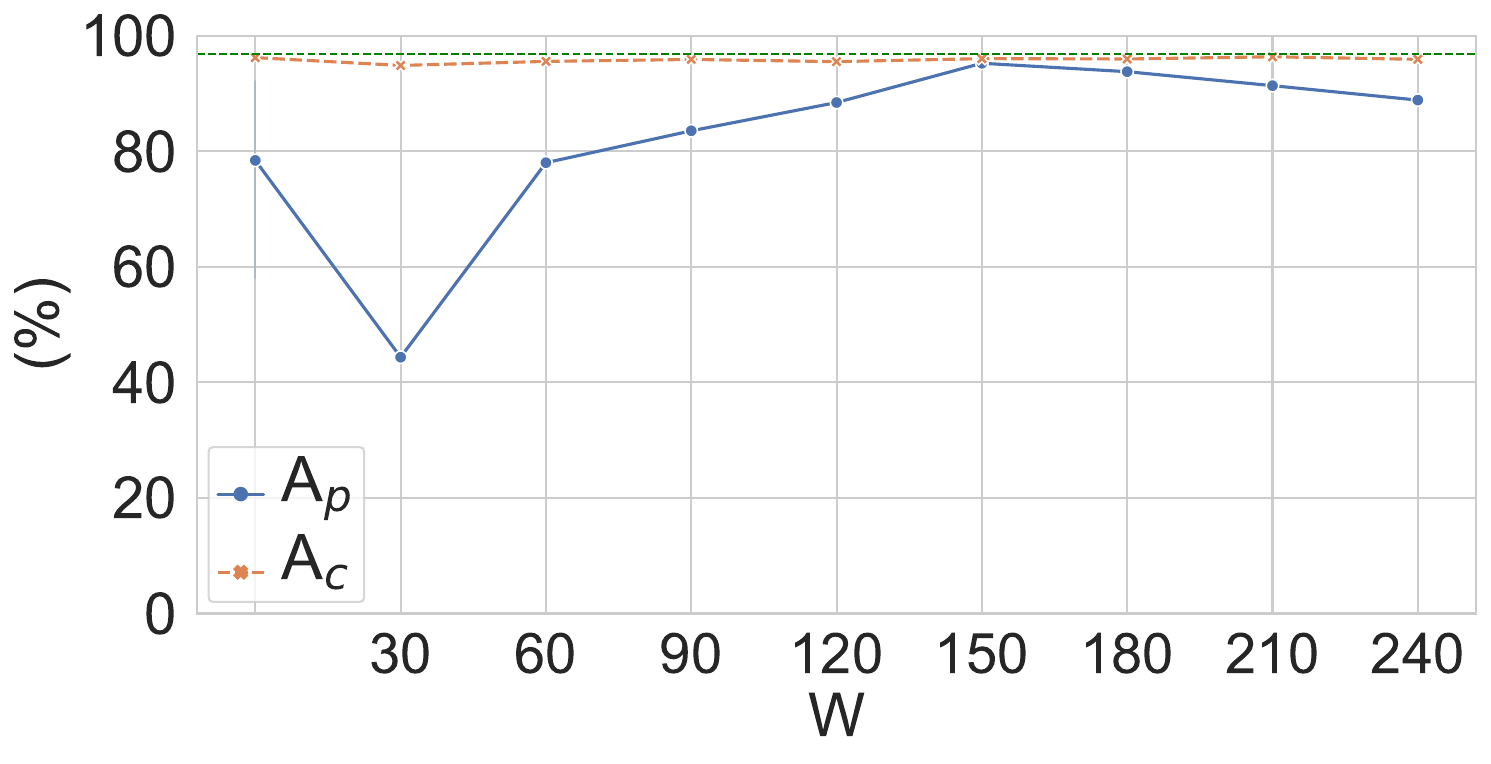}
    }
    \caption{The clean accuracy $A_c$ and attack success rate $A_p$ with respect to the value of the opacity parameter $W$.}
    \label{fig:W}
\end{figure}

\smallskip
\noindent
{\bf Optimizing the opacity parameter $W$.} 
We traverse the value range of the opacity parameter $W$, from 15 to 255 with the step of 15.
\figurename~\ref{fig:W} demonstrates how accuracy changes with respect to the opacity parameter $W$.
We find that the attack success rate $A_p$ of the red and green triggers 
does not monotonically change with respect to the opacity parameter $W$. 
We conjecture that this is because a transparent trigger (small $W$) 
is more likely to be overlooked by the model, 
while a trigger with high opacity (large $W$) {may influence the image semantics}. 
Specifically, the red, green, and blue triggers achieve the highest attack success rate $A_p$ 
when $W=90$, $W=60$, and $W=150$, respectively, 
which are set as the optimal values of the opacity parameter $W$.

\begin{figure}[t]
  \centering 
    \includegraphics[width=.48\textwidth]{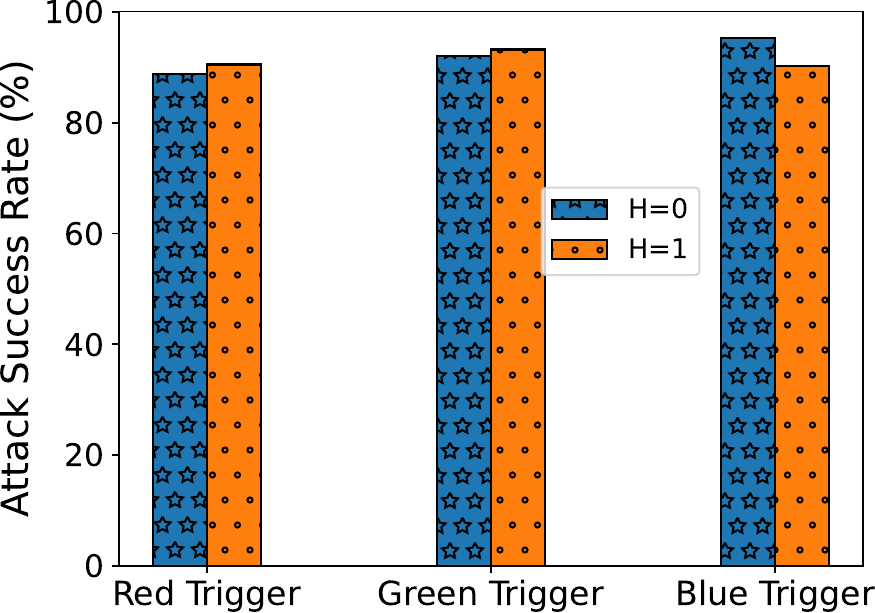}
    \caption{Attack success rate $A_p$ w.r.t. the value of the highlighting parameter $H$.}
    \label{fig:optimize-H} 
\end{figure}

\smallskip
\noindent
{\bf Optimizing the highlighting parameter $H$.} 
Similarly, we traverse the value range of the highlighting parameter $H$, i.e., $\{0,1\}$.
\figurename~\ref{fig:optimize-H} compares the attack success rate $A_p$ between when $H=0$ (trigger center not highlighted) 
and $H=1$ (trigger center highlighted). 
We find that the red and green triggers achieve a higher attack success rate $A_p$ when $H=1$, 
while the blue trigger achieves a higher attack success rate $A_p$ when $H=0$. 
{The difference is probably due to their different shapes}. 
Accordingly, we set $H=0$ for the blue trigger and $H=1$ for the other two types of triggers. 

\smallskip
\noindent {\bf Final Attack.} 
\tablename~\ref{tab:parameter-value} shows the final optimized value of each parameter. 
 \figurename~\ref{fig:optimize-benign}  and 
\figurename~\ref{fig:optimize-poison} 
show the clean accuracy and the attack success rate  of the baseline attack 
and the attack with iteratively optimized parameter in the order $L$, $K$, $W$, and $H$, respectively. 
We can observe that during the step-by-step optimization in the order $L$, $K$, $W$, and $H$, 
the attack success rate $A_p$ generally keeps increasing, 
while the clean accuracy $A_c$ remains stable and decreases no more than 0.4\% throughout the optimization process. 
Our final optimized attack achieves the attack success rate of 90.5\%, 93.2\%, and 95.3\% 
on the red, green, and blue triggers, respectively, significantly higher than the baseline attack. 
These demonstrate the effectiveness of the optimization approach and our attack \attackname.

\subsection{Many-to-One Attack}\label{sec:many-to-one-attack}
Previous experiments confirmed the effectiveness of \attackname when there is one designated trigger and one corresponding target label (i.e., one-to-one attack). Here, we demonstrate the capability of \attackname for many-to-one attack where the adversary specifies $k$ different triggers while associates them with the same target label. Without loss of generalizability, we set $k=2$. To train the backdoored model, we first randomly partition
the set of randomly selected clean images $T_{select}$ (5\% of the clean training dataset $T_{train}$) into two parts containing nearly identical number of samples, then add red trigger and green trigger to the two parts, respectively, 
and finally modify the labels of all poisoned images to ``stop and yield''. The trigger parameters are set as the optimal values found in \Cref{sec:exper-optimize}. The results are shown in \tablename~\ref{tab:many}. 
Although the red and green triggers
have the same shape,
the one-to-one attack has no more than 5\% attack success rate on the images with different trigger and target label from that involved in training. In contrast, although the poisoning rate is reduced from 5\% to 2.5\% for each physical trigger, the many-to-one attack is able to achieve 85.6\% and 89.2\% attack success rate on the red and green physical triggers, respectively. Also, compared to the one-to-one attack, the clean accuracy $A_c$ of the many-to-one attack only decreases by 0.8\%. 
These demonstrates the capability of \attackname in achieving many-to-one backdoor attacks. 

\subsection{Many-to-Many Attack}
We also demonstrate the capability of \attackname in launching the many-to-many backdoor attack where the adversary specifies $k$ different triggers each of which is associated with a distinct target label. We also set $k=2$, and the process of model training is the same as in \Cref{sec:many-to-one-attack} except that we set the ``stop and yield'' label and ``speed limit'' label for the red and green triggers, respectively. The results are shown in \tablename~\ref{tab:many}. 
The many-to-many attack achieves 84.9\% attack success rate on the green physical trigger, much higher than that of the one-to-one attack on ``green $\to$ $il100$''. Also, the clean accuracy $A_c$ of the many-to-many attack decreases only by 1.8\%, compared to the one-to-one attack. These demonstrate the capability of \attackname in achieving many-to-many backdoor attacks. However, 
the attack success rate of the many-to-many attack on the red trigger is slightly lower than that of the one-to-one attack on ``red $\to$ $ps$'', because the poisoning rate is reduced from 5\% to 2.5\%.

\begin{table}[t]
    \centering 
    \caption{Results of many-to-one and many-to-many attacks, where $ps$ and $il100$ are short for ``stop and yield'' and ``speed limi'' signs, respectively.} 
      \resizebox{1.\textwidth}{!}{
  \begin{tabular}{c|c|c|c|c}
    \hline
         \diagbox{\bf Train}{\bf Test} & {\bf $A_c$} & {\bf red $\to$ $ps$} & {\bf green $\to$ $ps$} & {\bf green $\to$ $il100$} \\ \hline
        \makecell[c]{{\bf One-to-one}  {\bf (red $\to$ $ps$})} & 96.8\% & 90.5\% & 5\% & 2.5\% \\ \hline
        \makecell[c]{{\bf One-to-one}  {\bf (green $\to$ $ps$})} & 96.4\% & 2\% & 93.2\% & 1.0\%  \\ \hline
        \makecell[c]{{\bf Many-to-one}  \\ {\bf (red $\to$ $ps$}) \& {\bf (green $\to$ $ps$})} & 96.0\% & 85.6\% & 89.2\% & N/A \\ \hline
        \makecell[c]{{\bf Many-to-many}  \\ {\bf (red $\to$ $ps$)} \& {\bf (green $\to$ $il100$)} } & 95.0\% & 80.3\% & N/A & 84.9\% \\ \hline
    \end{tabular}
    }
    \label{tab:many}
\end{table}

\subsection{Ablation Study}\label{sec:ablation_study}
In this subsection, we perform ablation experiments to further evaluate 
 the effectiveness of the proposed optimization method and the poisoned training dataset construction method by utilizing digital triggers to simulate the physical triggers.

\smallskip
\noindent
{\bf Swapping optimization order.} 
To validate the necessity of determining the {laser} parameter optimization order, 
we run the attack with an intentionally swapped optimization order. 
Specifically, we randomly swapped the optimization order of the location parameter $L$ and the scale parameter $K$, 
i.e. optimizing the attack with the order $K$, $L$, $W$, and $H$, 
instead of the order $L$, $K$, $W$, and $H$.  
As shown in \tablename~\ref{tab:final_accuracy_vari}, 
the attack success rate $A_p$ decreases on the red and green triggers, 
and the clean accuracy $A_c$ decreases on the red trigger. 
These demonstrate the significance of using appropriate optimization order 
and the effectiveness of the optimization order determining method used in our optimization approach.  
We notice that $A_p$ and $A_c$ remain unchanged for the blue backdoor 
because the optimized values of all parameters for this trigger do not change (cf.~\tablename~\ref{tab:final_para_veri}).

\begin{table}[t]
    \centering\setlength{\tabcolsep}{8pt}
     \begin{minipage}{0.98\textwidth}\centering
     \caption{The attack success rate $A_p$ and clean accuracy $A_c$ with the swapped optimization order.}
  \label{tab:final_accuracy_vari}
  \begin{tabular}{cccc}
  \toprule
  & Red & Green & Blue \\ \hline
  $A_p$ (original) & 90.5\% & 93.2\%  & 95.3\%\\
  $A_p$ (swapped) & 87.1\%$\downarrow$& 90.1\%$\downarrow$ & 95.3\%$-$\\
  $A_c$ (original) & 95.9\% & 96.3\%          & 96.0\% \\
  $A_c$ (swapped) & 95.8\%$\downarrow$& 96.3\%$-$          & 96.0\%$-$ \\
  \bottomrule
  \end{tabular}
    \end{minipage}
    \quad
     \begin{minipage}{0.98\textwidth}\centering
     \caption{Final optimized {laser} parameter with swapped optimization order. 
    Text in \red{red} highlights change compared to original order.}
    \label{tab:final_para_veri}
    \begin{tabular}{ccccc}
    \toprule
            &      $L$&      $K$&  	$W$&	        $H$\\\hline
       Red (original) &	$Random$& {6}&        90&    	{1}\\
       Red (swapped) &	$Random$& \red{5}&        90&    	\red{0}\\
       Green (original) &	$Random$& {6}&  {60}& 		{1}\\ 
       Green (swapped) &	$Random$& \red{7}&  \red{75}& 		\red{0}\\
       Blue (original) &	$Random$&       4&       150& 	         0\\
       Blue (swapped) &	$Random$&       4&       150& 	         0\\
    \bottomrule
    \end{tabular} 
    \end{minipage}
\end{table}

\smallskip
\noindent
{\bf Model transferability.} 
Previously, we assumed that the attacker can optimize the {laser parameters} 
on the target DNN model. 
To relax this assumption, we investigate the transferability of our approach 
by (1) applying the {laser parameter} values \emph{in each step} during the optimization process   
obtained on the ResNet model to 
ViT~\cite{ViT} 
and GoogLeNet~\cite{GoogLeNet} models. 
The results are reported in \tablename~\ref{tab:vit} and \tablename~\ref{tab:googlenet}. 
Similar to the ResNet model, 
the attack success rate $A_p$ keeps increasing on at least one trigger, and finally, the optimized attack achieves a much higher attack success rate $A_p$ than the baseline attack; 
{(2) applying the \emph{final} optimized laser parameters obtained on ResNet to Yolo-V8~\cite{yolov8}, 
achieving the attack success rate of 83.2\%, 88.2\%, and 96.54\% with the red, green, and blue triggers, respectively.}
These results confirm the generability of the optimized {laser parameter}, 
the model transferability of our laser parameters optimization approach, 
{and our attack's effectiveness across different models.}

\begin{table}[t]
    \centering\setlength{\tabcolsep}{8pt}
    \begin{minipage}{0.98\textwidth}\centering
     \caption{Attack success rate $A_p$ of the victim model ViT at each step during the optimization process.}
    \begin{tabular}{cccc}
    \toprule
        ~ & Red & Green & Blue  \\ \hline
        Baseline & 31.3\% & 52.3\% & 58.1\%  \\ 
        L & 62.4\%$\uparrow$ & 87.6\%$\uparrow$ & 63.2\%$\uparrow$  \\ 
        K & 62.4\%$-$ & 87.6\%$-$ & 98.1\%$\uparrow$  \\ 
        W & 62.4\%$-$ & 88.1\%$\uparrow$ & 99.5\%$\uparrow$  \\ 
        H  & 70.7\%$\uparrow$ & 90.3\%$\uparrow$ & 99.5\%$-$ \\ 
        \bottomrule
    \end{tabular}
    \label{tab:vit}
    \end{minipage}
  
    \begin{minipage}{0.98\textwidth}\centering
          \caption{Attack success rate $A_p$ of GoogLeNet 
          at each step during laser parameters optimization.}
   \begin{tabular}{cccc}
    \toprule
      ~ & Red & Green & Blue  \\ \hline
      Baseline & 40.7\% & 48.1\% & 57.2\%  \\ 
      L & 81.5\%$\uparrow$ & 83.8\%$\uparrow$ & 70.9\%$\uparrow$  \\ 
      K & 81.5\%$-$ & 83.8\%$-$ & 76.4\%$\uparrow$  \\ 
      W & 81.5\%$-$ & 85.3\%$\uparrow$ & 89.8\%$\uparrow$  \\ 
      H & 85.3\%$\uparrow$ & 88.8\%$\uparrow$ & 89.8\%$-$ \\ 
      \bottomrule
  \end{tabular}
  \label{tab:googlenet}
    \end{minipage}
\end{table}

\begin{figure}[t]
    \centering
    \subfloat[Red trigger]{
    \includegraphics[width=0.42\textwidth]{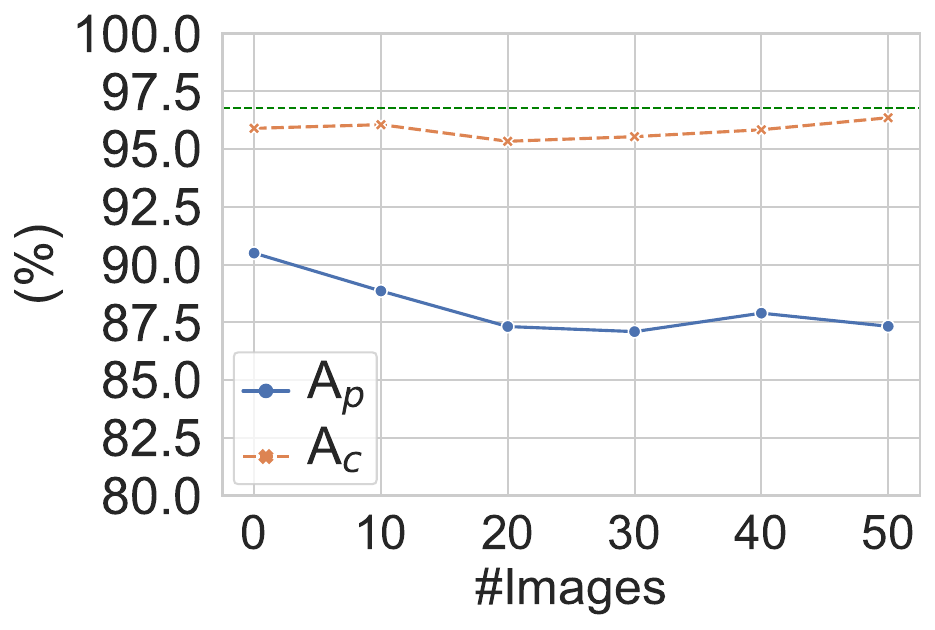}
    }
      \subfloat[Green trigger]{
    \includegraphics[width=0.42\textwidth]{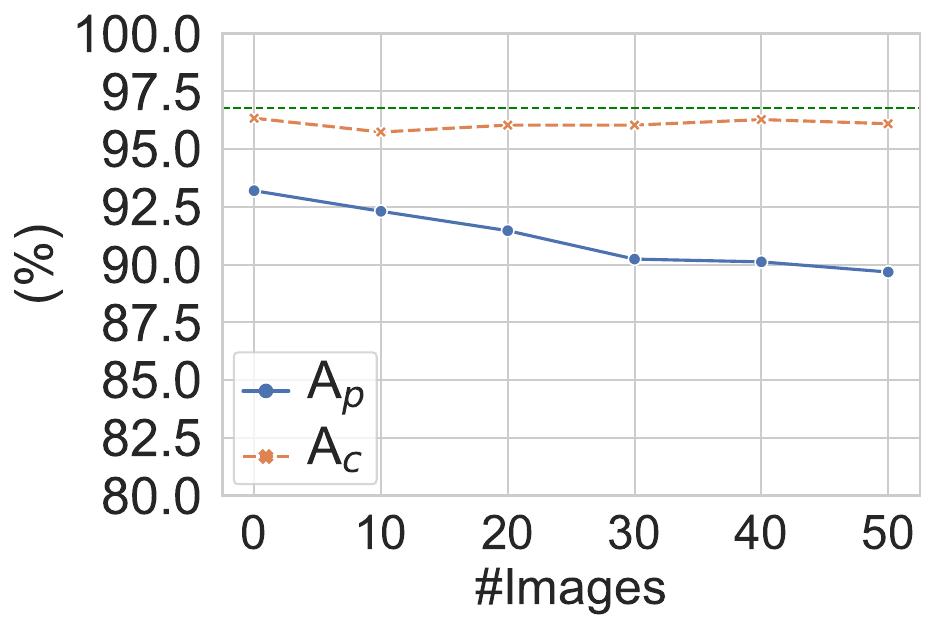}
    }
    
      \subfloat[Blue trigger]{
    \includegraphics[width=0.42\textwidth]{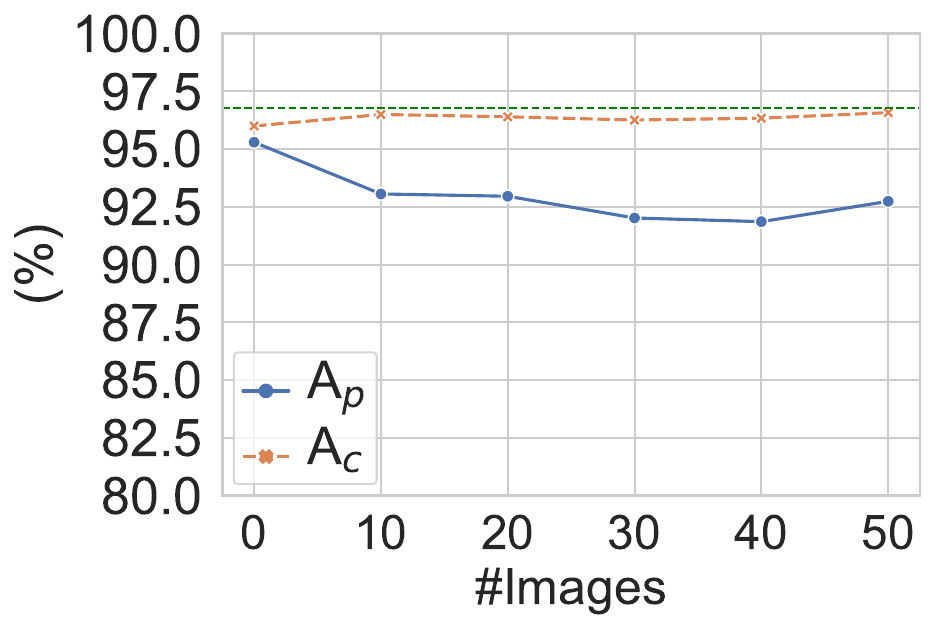}
    }
    \caption{The clean accuracy $A_c$ and attack success rate $A_p$ with respect to the number of poisoned training images with the physical triggers. \#Images=0 denotes only using digital triggers when constructing the poisoned training dataset.}
    \label{fig:change_digit_to_physical}
\end{figure}

\smallskip
\noindent
{\bf Replacing digital triggers with physical ones.} 
To reduce attack overhead, we simulate physical triggers with digital ones for constructing poisoned training datasets. Here we compare the attack effectiveness between using digital triggers alone and including physical triggers. For each trigger, we replace the digital trigger with its physical counterpart in $M$ randomly selected images from $P_{train}$, with $M$ ranging from 10 to 50. As shown in \figurename~\ref{fig:change_digit_to_physical}, attacks using physical triggers in the training dataset consistently achieve lower success rates than those using only digital triggers, likely because digital triggers can simulate a broader range of physical triggers due to easier parameter control. This demonstrates the effectiveness of utilizing digital triggers to simulate the physical triggers when constructing the poisoned training dataset.

\section{Discussion}\label{sec:discussion}

\subsection{Failure Analysis}\label{sec:failure}
\begin{figure*}
    \centering
    \begin{minipage}[t]{0.48\textwidth}
        \centering
        \includegraphics[width=0.4\textwidth]{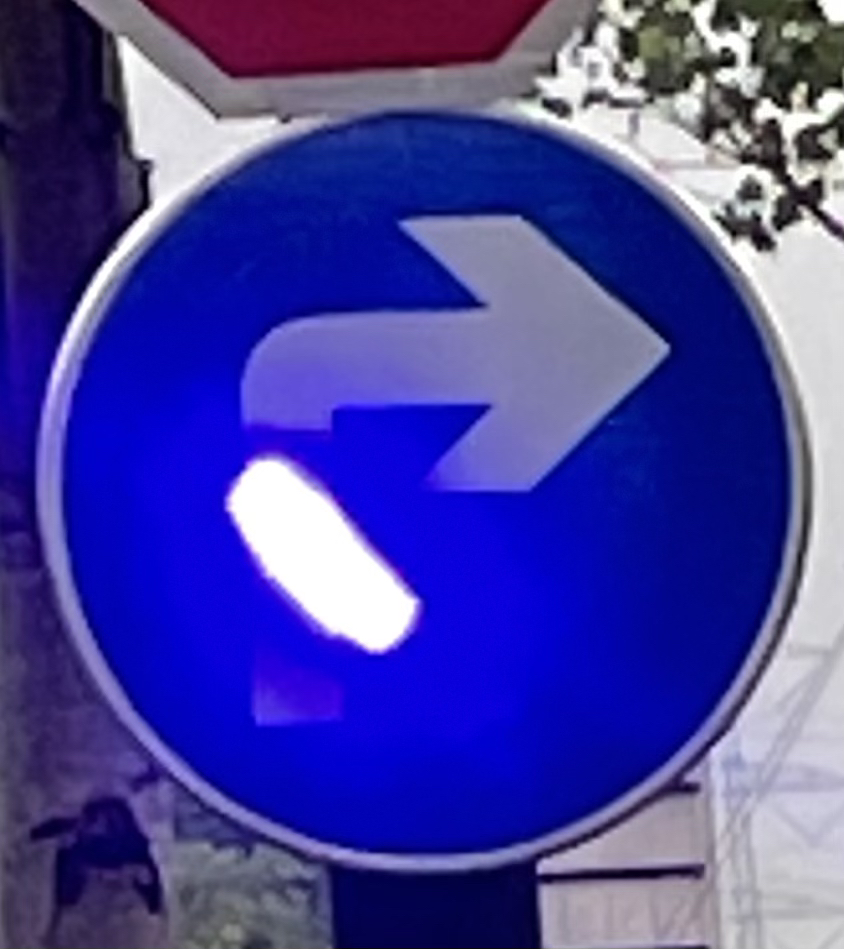}
        \includegraphics[width=0.45\textwidth]{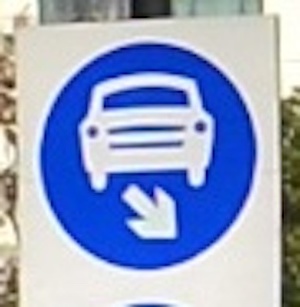}
    \caption{Trigger changing semantics}
    \label{fig:fail1}
\end{minipage}
\quad
    \begin{minipage}[t]{0.48\textwidth}
        \centering
            \includegraphics[width=0.45\textwidth]{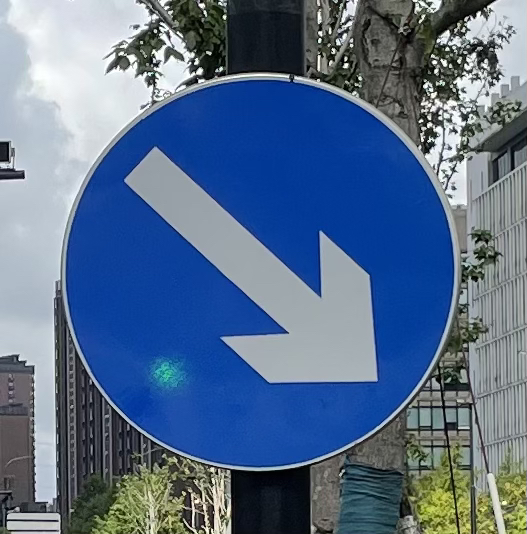}
            \includegraphics[width=0.47\textwidth]{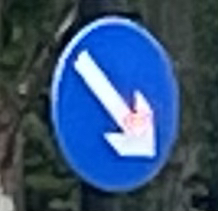}
    \caption{Bad trigger quality}
    \label{fig:fail2}
    \end{minipage}
\end{figure*}

While our attack can achieve high success rate, 
there are still few cases on which \attackname fails. 
After analyzing the failed cases, 
we find that the failures are mainly due to the following two reasons:
(1) {\bf Triggers changing semantics.} 
We notice that sometimes an embedded trigger can change the image semantics 
from the original traffic sign to another traffic sign. 
For instance,
an overexposed blue trigger, e.g., \figurename~\ref{fig:fail1} (left), looks like a slanted arrow in the ``Motor vehicle lane'' sign as shown in \figurename~\ref{fig:fail1} (right). 
As a result, the victim model will classify it to the ``Motor vehicle lane'' sign rather than our target label. 
(2) {\bf Bad trigger quality.} 
When a trigger is too bright or too dim, e.g., as demonstrated in \figurename~\ref{fig:fail2}, 
they got unnoticed by the victim model, leading to a failed attack.

\subsection{Countermeasure}\label{sec:defense}
{We consider five defenses. The former four from prior works 
were shown promising and the last one is our designed adaptive defense. 
While they vary from backdoor detection~\cite{NC} to poison data detection~\cite{AC, SS}, run-time trigger detection~\cite{STRIP} and removal, \attackname shows robustness against all of them, urging for more effective defenses to mitigate laser physical backdoor attacks.}

\smallskip \noindent {\bf Neural Cleanse (NC)}~\cite{NC}
{deems a model backdoored if a label requires much smaller norm of modifications to classify all examples into it. Applied to our attack, 95\% of unaffected labels have smaller norms than the target label, indicating NC's unreliability. This is due to the larger laser spot size compared to sticker-based triggers~\cite{gu2017badnets}, resulting in a higher norm for the target label.}

\smallskip \noindent {\bf Activation Clustering (AC)}~\cite{AC}
{assumes that inputs with triggers activate distinct neurons from clean inputs, detecting poisoned training data by comparing activation patterns. Applied to \attackname, AC yields a 90\% false negative rate and 29\% false positive rate, likely due to significant overlap in neuron activation between inputs with laser-based triggers and clean inputs.}

\smallskip \noindent {\bf Spectral Signatures (SS)}~\cite{SS}
{observes that inputs with triggers exhibit differences in the covariance spectrum of neural network feature representations compared to clean inputs, using statistical methods like SVD to detect outliers (poisoned data). However, it identifies only 8\% of poisoned samples for \attackname, indicating that our attack remains a potent real-world threat.}

\smallskip \noindent {\bf STRIP}~\cite{STRIP}
{blends an inference input with random clean inputs and flags it as containing triggers if its output randomness (entropy) is below a threshold (set for a 5\% false positive rate per \cite{wenger2021backdoor}). Applied to \attackname, STRIP fails to detect 63\% of triggered inputs, likely due to the diversity of physical laser-based triggers, which maintain high output randomness even after blending.}

\smallskip \noindent {\bf Laser Removal (LR)}
{uses prior knowledge of the laser spot's color and shape to mask matching regions and replace them with pixel interpolation.  
The attack success rate is reduced but still relatively high (over 48\%), 
since inconsistencies in shape and color caused by extreme projection angles and lighting distortions leave many spots undetected. 
Additionally, misidentifying traffic sign areas as spots reduces normal accuracy by 14\%. These indicate that it is challenging to mitigate physical triggers without affecting normal accuracy~\cite{li2020light}.}
\section{Conclusion}\label{sec:concl}
In this work, we designed laser-based physical triggers and proposed \attackname, a novel laser-driven backdoor attack targeting deep neural networks. Leveraging lasers' long-distance transmission and instant imaging capabilities, \attackname achieves remote control, high temporal stealthiness, and superior flexibility and mobility, addressing the limitations of existing backdoor methods. 
To enhance attack performance, we introduced an effective parameter selection method for laser-based triggers. Experiments on traffic sign recognition models, using both open-source and our collected real-world datasets, demonstrate the high effectiveness of \attackname, negligible impact on normal inputs, and the effectiveness of the optimization approach. 
Our work sheds light on more practical and feasible physical backdoor attacks 
and contributes the first traffic sign dataset featuring physical laser triggers to advance research in this domain.

\begin{credits}
\subsubsection{\ackname}
We thank our anonymous shepherd and the reviewers for their constructive feedback and suggestions. 
This research is partially supported by CAS Project for Young Scientists in Basic Research (Grant No. YSBR-040), ISCAS New Cultivation Project (Grant No. ISCAS-PYFX-202201), and ISCAS Basic Research (Grant No. ISCAS-JCZD-202302). 

\subsubsection{\discintname}
The authors have no competing interests to declare that are
relevant to the content of this article.
\end{credits}

\bibliographystyle{splncs04}
\bibliography{reference}

\end{document}